\documentclass{article}
\usepackage[margin=1in]{geometry}

\usepackage{graphicx}

\usepackage{amsmath}
\usepackage{amssymb}
\usepackage{TeXSanity}
\usepackage{titling}
\usepackage{lipsum}
\usepackage{booktabs}
\usepackage{soul}
\usepackage{color}

\usepackage{fancyhdr} % for distro statement
\fancypagestyle{distro-statement}
{
    \fancyfoot[C]{\textbf{Distribution Statement A.}  Approved for public release: distribution is unlimited.}
}

\usepackage{setspace}

\usepackage{caption}
\usepackage{subcaption}
\usepackage{wrapfig}

\usepackage{tikz}
\usetikzlibrary{shapes,arrows}

\usepackage[normalem]{ulem} % fixed bib running into margins
\usepackage{authblk} % for authors and affiliations
\graphicspath{{figures/}}
\usepackage[section]{placeins} % force figures to reside in section they are delcared

\begin{document}

\title{Mapping Microstructure: Manifold Construction for Accelerated Materials Exploration}

\author[1,2]{Simon A. Mason}
\author[3]{Megna N. Shah}
\author[3]{Jeffrey P. Simmons}
\author[4]{Dennis M. Dimiduk}
\author[*,1,5]{Stephen R. Niezgoda}

\affil[1]{Department of Materials Science and Engineering, The Ohio State University, 2041 College Road, Columbus, OH 43210, USA}
\affil[2]{US Naval Research Laboratory, 4555 Overlook Ave. SW, Washington, DC 20375, USA}
\affil[3]{Materials and Manufacturing Directorate, Air Force Research Laboratory, Wright-Patterson AFB, OH 45433, USA}
\affil[4]{Blue Quartz Software, LLC, 400 S. Pioneer Blvd, Springboro, OH 45066, USA}
\affil[5]{Department of Mechanical and Aerospace Engineering, The Ohio State University, 2041 College Road, Columbus, OH 43210, USA}
\affil[*]{Corresponding author: niezgoda.6@osu.edu}
\date{September 2025}

\maketitle

\begin{abstract}
Accelerating materials development requires quantitative linkages between processing, microstructure, and properties. In this work, we introduce a framework for mapping microstructure onto a low-dimensional material manifold that is parametrized by processing conditions. A key innovation is treating microstructure as a stochastic process, defined as a distribution of microstructural instances rather than a single image, enabling the extraction of material state descriptors that capture the essential process-dependent features. We leverage the manifold hypothesis to assert that microstructural outcomes lie on a low-dimensional latent space controlled by only a few parameters. Using phase-field simulations of spinodal decomposition as a model material system, we compare multiple microstructure descriptors (two-point statistics, chord-length distributions, and persistent homology) in terms of two criteria: (1) intrinsic dimensionality of the latent space, and (2) invertibility of the processing-to-structure mapping. The results demonstrate that distribution-based descriptors can recover a two-dimensional latent structure aligned with the true processing parameters, yielding an \textit{invertible and physically interpretable mapping} between processing and microstructure. In contrast, descriptors that do not account for microstructure variability either overestimate dimensionality or lose predictive fidelity. The constructed material manifold is shown to be locally continuous, wherein small changes in process variables correspond to smooth changes in microstructure descriptors. This data-driven manifold mapping approach provides a quantitative foundation for microstructure-informed process design and paves the way toward closed-loop optimization of processing--structure--property relationships in an integrated materials engineering context.
\end{abstract}

% \keywords{Microstructure Mapping, Material Manifold, Processing--Structure Relationship, Stochastic Representation, Manifold Learning}

\thispagestyle{distro-statement}

%%%%%%%%%%%%%%%%%%%%%%%%%%%%%%%%%%%%%%%%%%%%%%%%%%%%%%%%%%%%
% \setcounter{page}{1}

\clearpage
\pagenumbering{arabic}
% \doublespacing

%%%%%%%%%%%%%%%%%%%%%%%%%%%%%%%%%%%%%%%%%%%%%%%%%%%%%%%%%%%%
\section{Introduction}
\label{intro}

% from reviewer #2: The main contributions should be stated more sharply early on (stochastic microstructure representation, dimensionality + invertibility criteria, descriptor comparison).

Accelerated materials exploration hinges on a data-driven mapping between processing conditions and the resulting microstructure that is both low-dimensional and invertible. In this work, we make three related contributions: first, we introduce a stochastic representation of microstructure in which the microstructure itself is treated as a random variable governed by the processing protocol rather than a static image; second, we formulate quantitative dimensionality and invertibility criteria based on manifold theory to assess whether a descriptor captures the intrinsic degrees of freedom of the process and can be uniquely mapped back to it; third, we apply these criteria to compare several candidate descriptors and determine which best preserves the controllable features of the underlying process. 

A key strategy for successful technology development as proposed by the National Research Council is through “quick, iterative development cycles”, with quantifiable measurements providing feedback information on the impact of developmental changes \cite{AcceleratingTechnologyTransition2004}. In order to enable these rapid cycles, data-driven practices and quantitative analysis can be used to formalize the relationships between processing, structure, properties, and performance, defining a mapping between these domains --- termed the \textit{material manifold}.

Modern materials engineering has evolved from ad‑hoc metallurgy to Integrated Computational Materials Engineering (ICME), where microstructure‑sensitive design (MSD) and microstructure informatics (MI) aim to accelerate the processing–structure–property cycle. A central challenge is navigating this high‑dimensional space efficiently: given a target property, we must predict how to adjust processing parameters to attain the desired microstructure --- a concept we term \textit{navigability} on the material manifold. We hypothesize that the stochastic representation of microstructure yields a low-dimensional manifold, mapping the relationship between material processing conditions and resulting microstructure. By applying our dimensionality and invertibility criteria to various descriptors, we test this hypothesis and provide a principled basis for rapid, iterative design cycles in materials development.

\section{Background}

\subsection{Microstructure as a Stochastic Process}

\begin{figure}[!h] \begin{centering} \includegraphics[width=0.8\linewidth]{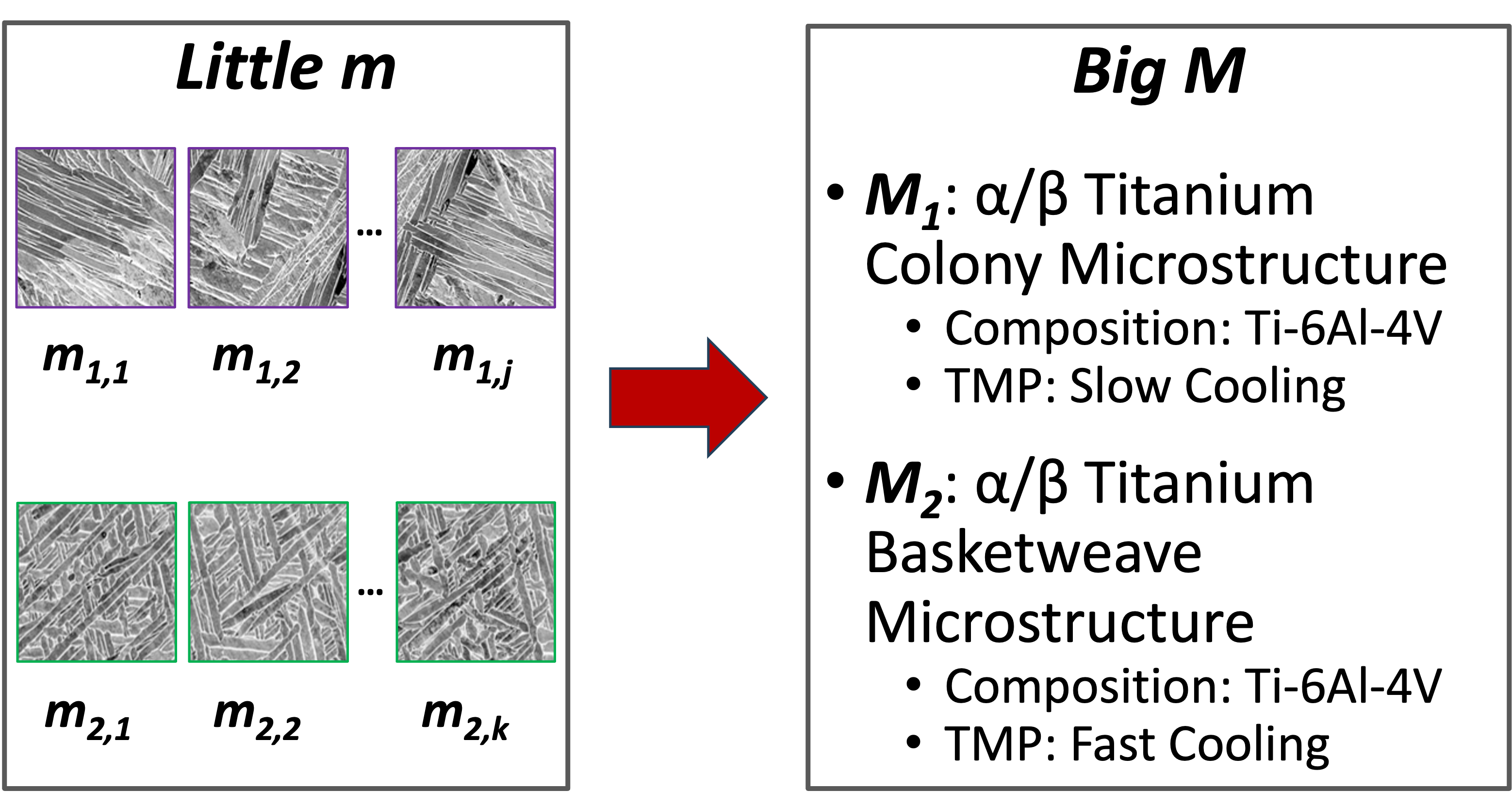} \caption{Microstructure as a stochastic process.} \label{fig:stoch_process} \end{centering} \end{figure}

A foundational concept for understanding microstructure's role in PSPRs is treating microstructure as a stochastic process \cite{niezgodaStochasticRepresentationMicrostructure2010}. In this definition, microstructure is a probability distribution of ``essentially equivalent'' microstructure realizations sampled from the stochastic process. This concept, shown in \ref{fig:stoch_process}, is generalized in this work to instead define the \textit{material state} as a stochastic process - with microstructure being one descriptor set to represent material state information. We denote the stochastic process by `big $M$' (the `$M$-state') and individual realizations within one $M$-state by `little $m$' (an `$m$-instance'). The distinction is that an $m$‑instance is a microstructural image or its quantitative description, whereas an $M$-state is the statistical ensemble of all such instances. The term ``microstructure'' is often used to refer to an image, or a quantitative descriptor set derived from an image. From here, we use the term ``microstructure'' to mean the set of $m$-instances used to describe a $M$-state. The $M/m$ concept is not novel to this work, however it has mainly been used to describe distributions of $m$-instances within one $M$-state or to compare discrete classifications of $M$-states \cite{niezgodaStochasticRepresentationMicrostructure2010}. This work will expand it to allow for comparisons of a continuum of $M$-states across the material manifold.

\subsection{Processing-Microstructure-Property Relationship and the Need for Multi-scale Representations}

\begin{figure}[!h] \begin{centering} \includegraphics[width=0.5\linewidth]{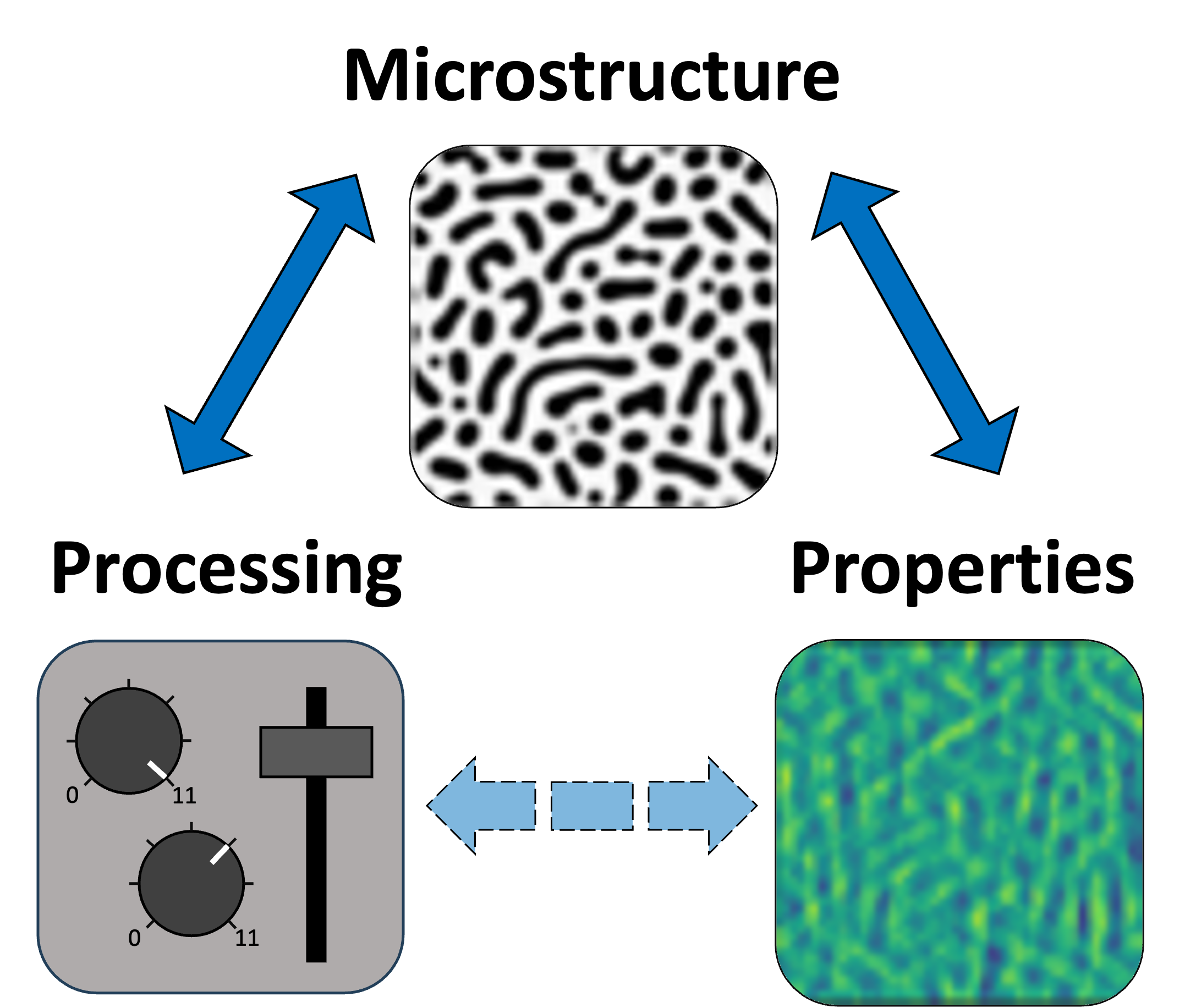} \caption{Processing-structure-property relationship triangle.} \label{fig:pspr_traingle} \end{centering} \end{figure}

The processing-structure-property relationship is a fundamental concept in materials science. Being able to harness the linkage between material processing, its impact on the resulting microstructure, and the property response of these materials, is a powerful tool for MSD \cite{fullwoodMicrostructureSensitiveDesign2010}. Under this methodology, the material domain, design space, and resultant properties are explored with respect to the microstructure. The relationship between processing and properties is difficult to determine directly but can be understood by relating processing to microstructure and microstructure to properties \cite{shimEffectsProcessParameters2021}, as shown in Figure \ref{fig:pspr_traingle}. The processing-structure-property relationship requires a complete description from the microstructure to be able to understand how properties relate to processing conditions.

As of today, this ``microstructure bridge'' is a primarily academic construct, but increasing complexities in the processing domains and material performance requirements necessitate the development of tools to explore their relationships to microstructure. In practical engineering and historical contexts, materials were explored, developed, and defined by their properties with no insight into microstructure, either because the characterization tools did not exist, or the quantification methods did not allow for microstructure descriptors to be integrated with engineering design. The quantitative representation of microstructure allows for materials state descriptors to be \textit{explicitly} defined on characterizations of microstructure features.

\subsection{Quantitative Representation of Microstructure} 

Microstructure is described by Fullwood et al. as the ``myriad features of internal structure of heterogeneous materials across length scales ranging from the atomic through the mesoscale to approaching the macroscale'' \cite{fullwoodMicrostructureSensitiveDesign2010}. \footnote{This description aligns with the current teachings of MSE, but significantly differs from that of the Wikipedia definition.} As the prefix ``micro'' indicates, we are often focused on the microscopic structure; however, modern characterization tools permit characterization at even sub-atomic scales, and in the case of components having compositional or processing gradients, important microstructural features can extend well beyond the millimeter length scale. The entire range can be understood as ``microstructure''. Microstructure quantification and reduced-order representations of microstructure are needed to understand PSPRs \cite{bugas_grain_2024}.

Three commonly used approaches for the quantitative representation of microstructure (QRM) are: (1) the explicit calculation of feature distributions \cite{wodoMicrostructuralInformaticsAccelerating2016}, \cite{groeberFrameworkAutomatedAnalysis2008}, \cite{groeberFrameworkAutomatedAnalysis2008a}, (2) the implicit representation of features using reduced-order models \cite{torquatoSTATISTICALDESCRIPTIONMICROSTRUCTURES2002}, \cite{kalidindiMicrostructureInformaticsUsing2011}, \cite{dlotko2016topological}, and (3) machine learning featurization \cite{decostComputerVisionApproach2015}, \cite{lenau2024physics}, \cite{muellerMachineLearningMaterials2016}. At their core, all three methods share a common approach, calculating reduced-order representations of microstructural features, just with different levels of human interaction. This work focuses on techniques falling into the second category as a method for extracting QRMs from microstructure images.

\section{The Material Manifold}
\label{manifold}

The fundamental contribution of this work revolves around the construction of a material state manifold, or a low-dimensional domain on which each point represents a unique material state. The material manifold is both a conceptual framework to visualize and analyze the variety of material states that exist in a material system \cite{decostExploringMicrostructureManifold2017}, as well as a mathematical object that establishes properties needed for iterative mappings between process and structure \cite{schoeneman2018entropy}, \cite{wodoMicrostructuralInformaticsAccelerating2016}. The manifold itself represents a complete set of all possible material states spanning the design space for a given range of composition and processing conditions. While the concept of a manifold embedding has already been adopted into materials science and machine learning \cite{sundararaghavanMicrostructuralStateVariables2023}, \cite{abelExploringManifoldNeural2024}, this work formalizes the requirements of quantitative descriptors needed to be applied to this type of analysis.

\subsection{Defining Microstructure as a Function of Processing}
\label{sec:funcProc}

Colloquially, materials scientists are known to say phrases like ``microstructure is a function of processing''. In our context, this means that the material state is a function of processing, and that microstructure information derived from characterization methods are indicators or descriptors of the material state. Furthermore, characterization datasets or images are often analyzed so that they become formal quantitative descriptors of the material state, as demonstrated later in this writing.

To formalize this notion, $M$ will be used to represent the material state domain and $\Theta$ will be used to represent the processing domain. The forward mapping, $f: \Theta \rightarrow M$, describes microstructure as a function of processing. The inverse mapping, $g: M \rightarrow \Theta$, describes the recovery of parameters as part of a feedback loop, where knowledge of $M$ permits one to know $\Theta$. Figure \ref{fig:tikz_simple} shows a diagram of the feedback loop.%, with one complete cycle, from parameters to material state and back, consisting of $g(f(\Theta))=\Theta$. 

\tikzstyle{block} = [draw, fill=blue!20, rectangle, minimum height=3em, minimum width=6em] \tikzstyle{sum} = [draw, fill=blue!20, circle, node distance=1cm] \tikzstyle{input} = [coordinate] \tikzstyle{output} = [coordinate] \tikzstyle{pinstyle} = [pin edge={to-,thin,black}] \tikzstyle{terminator} = [rectangle, draw, text centered, rounded corners, minimum height=2em] \tikzstyle{process} = [rectangle, draw, text centered, minimum height=2em] \tikzstyle{decision} = [diamond, draw, text centered, minimum height=2em] \tikzstyle{data}=[trapezium, draw, text centered, trapezium left angle=60, trapezium right angle=120, minimum height=2em] \tikzstyle{connector} = [draw, -latex'] \tikzstyle{point} = [coordinate]

\begin{center}
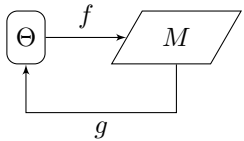
 \begin{tikzpicture}[auto,node distance=2cm,>=latex']

\node [terminator] (param) {$\Theta$}; \node [data, right of=param] (bigM) {$M$}; \node [point, below of=param,node distance=1cm] (pt1) {}; \node [point, below of=bigM,node distance=1cm] (pt2) {};

\draw [->] (param) -- node {$f$} (bigM); \draw[->] (bigM.south) -- (pt2) -- node {$g$} (pt1) -- (param.south);

\end{tikzpicture} \captionof{figure}{Invertible mapping between the processing and structure domains.} \label{fig:tikz_simple} \end{center}

While there are many potential ways to parameterize material states, we choose to do so based on the processing conditions, separating $\Theta$ as our \textit{controllable} domain, and $M$ as our \textit{observable} domain. This parameterization assumes a bijective mapping between process and microstructure, where there is a one-to-one correspondence between processing conditions and material states. For the subset of phase field simulations used in this work, a bijective mapping holds. This restriction is necessary for treating materials development as an inverse problem, where we are able to reduce the forward model from a many-to-one relationship to a one-to-one relationship through the stochastic representation of microstructure.

%%%%%%%%%%%%%%%%%%%%%%%%%%%%%%%%%%%%%%%%%%%%%%%%%%%%%%%%%%%%

\subsection{The Material Manifold}

The material manifold, $\mathcal{M}$, is parameterized by $\Theta$, the region of all possible controlling parameters of the material system (i.e. composition and process parameters), where $\theta$ represents a unique set of processing parameters and $M^{\theta}$ is the material state generated from that process, as shown in Equation \ref{eqn:manifold}. This definition hinges on the prior interpretation that different processing conditions result in different material states \cite{kurzSolidificationMicrostructureProcessingMaps2001}.

\begin{equation}
\label{eqn:manifold}
\mathcal{M} = \{ M^{\theta}: \theta \in \Theta \}
\end{equation}

\begin{figure}[ht] \begin{centering} \includegraphics[width=0.8\linewidth]{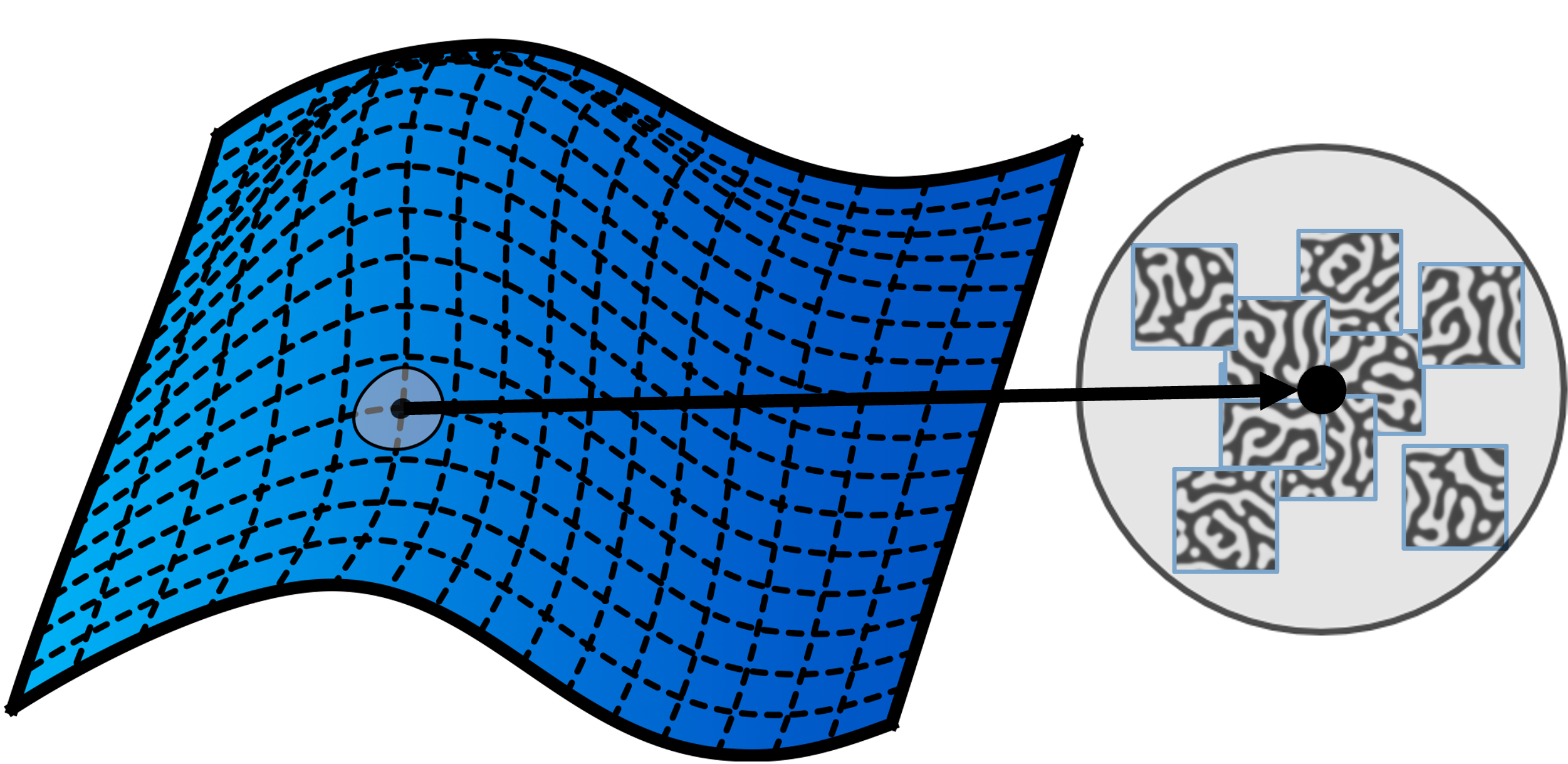} \caption{Schematic manifold and distribution of $m$-instances within a $M^{\theta}$-state} \label{fig:mani_sample} \end{centering} \end{figure}

The manifold consists of $m$-instances generated from a random process defined over $\Theta$ and parameterized by $\theta=(\alpha,\beta)$, where $\alpha$ and $\beta$ are generic parameters. $m$-instances are realizations from the $M^\theta$-state having unique spatial/deterministic information resulting from the stochastic nature of the generating process (i.e. thermal fluctuations). Figure \ref{fig:mani_sample} illustrates a unique $M^{\theta}$-state sampled from the surface of a generic manifold and the corresponding set of $m$-instances that exists within the distribution of $M^{\theta}$. The micrographs used in this section are spinodal decomposition microstructures, further described in Section \ref{sec:pf_system}.

\begin{figure}[!htb] \begin{centering} \includegraphics[width=0.8\linewidth]{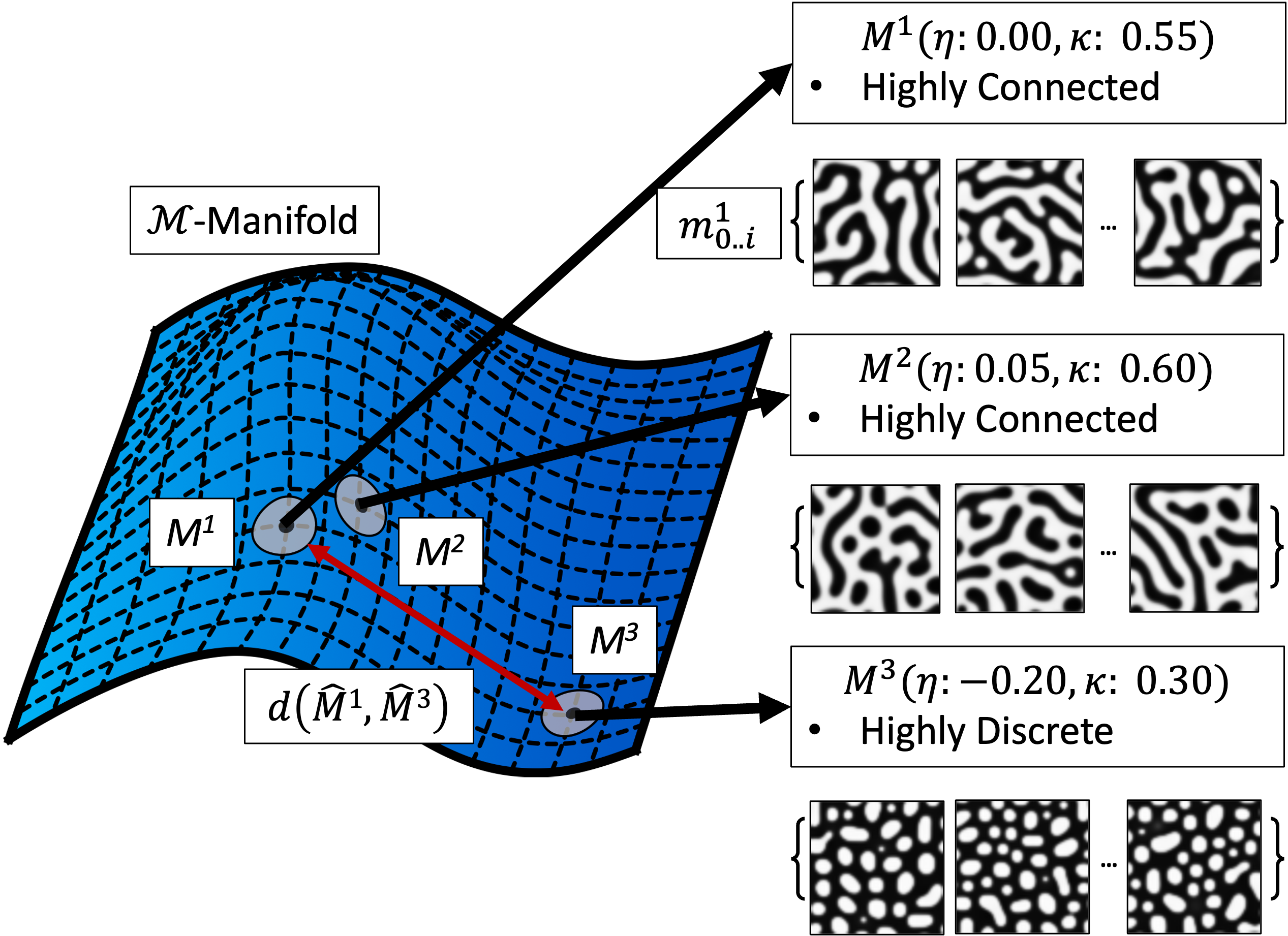} \caption{Schematic manifold and $M^{\theta}$-states with corresponding sets of $m$-instances.} \label{fig:mani_3M_classes} \end{centering} \end{figure}

Figure \ref{fig:mani_3M_classes} shows a $\mathcal{M}$-manifold highlighting three different $M^{\theta}$-states: two represented by similar microstructures that exist close to each other on the manifold ($M^1$, $M^2$), and one having different microstructures further away on the manifold ($M^3$). Building on the metrizable nature of the manifold, a notion of similarity and dissimilarity between $M^{\theta}$-states can be established.

\subsection{The Stochastic Representation of Microstructure and Microstructure Distance Metrics}

In understanding microstructure as a stochastic process, $m$-instances are sampled from $M$-state distributions. Microstructure descriptors can be established for which equivalent microstructures have equivalent descriptors, within some acceptable tolerance. These microstructure descriptors serve to capture the stochasticity of the generating process, detailing $M$-state information from sets of $m$-instances. For any arbitrary microstructure descriptor, a distance metric can be computed to calculate the degree of similarity between $m$-instances, or additionally the distances between distributions of $M$-states. In this section, analysis will be shown using the two-point correlation function as a descriptor due to its relative familiarity in materials science \cite{torquatoSTATISTICALDESCRIPTIONMICROSTRUCTURES2002}. Specifics on how calculations were completed will be shown in the later Section \ref{sec:microDescriptor}.

In order to learn information on $M$-states, microstructure descriptors, $D$, can be calculated on individual $m$-instances to return descriptor vectors upon which distance calculations can be performed. Various approaches to calculating microstructure distance metrics have been explored \cite{simmons2019statistical}, \cite{callahan2019distance}, however this work defines the metric descriptor distance (MDD), $d$, as the p-Minkowski norm between descriptor vectors, shown in Equation \ref{eq:mdd}. For this work, $p=1$, following the analysis done by Sundararaghavan et al. \cite{sundararaghavanMicrostructuralStateVariables2023} showing the tendency of higher $p$ values to over-estimate the dimensionality of systems.

\begin{equation} \label{eq:mdd} d(m^1, m^2) = (\sum_i{|D^1_i - D^2_i|^p})^{1/p} \end{equation}

\begin{figure}[!ht] \centering \begin{subfigure}[c]{0.50\textwidth} \includegraphics[width=\textwidth]{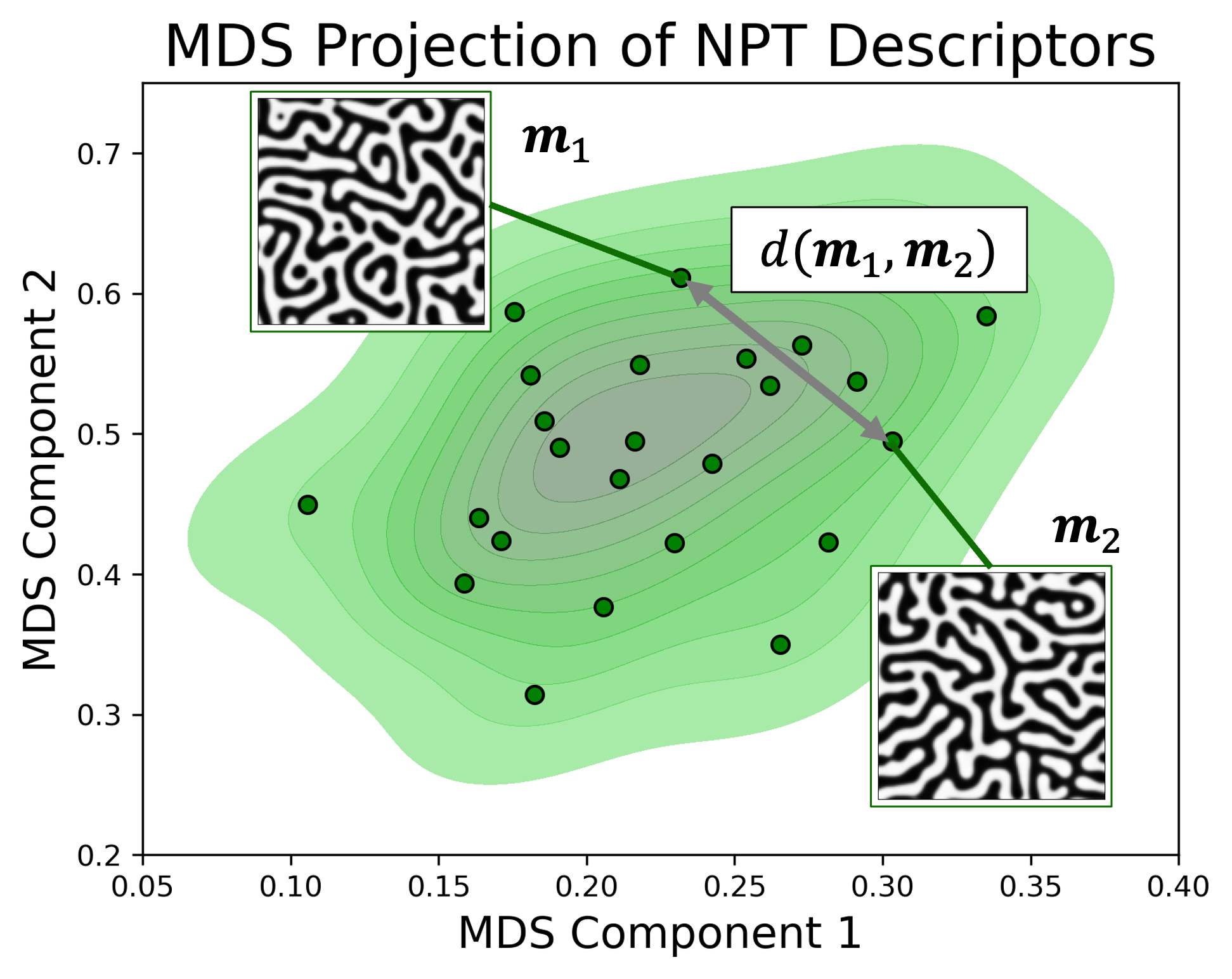} \caption{} \label{fig:mds_kde} \end{subfigure} \hfill \begin{subfigure}[c]{0.45\textwidth} \includegraphics[width=\textwidth]{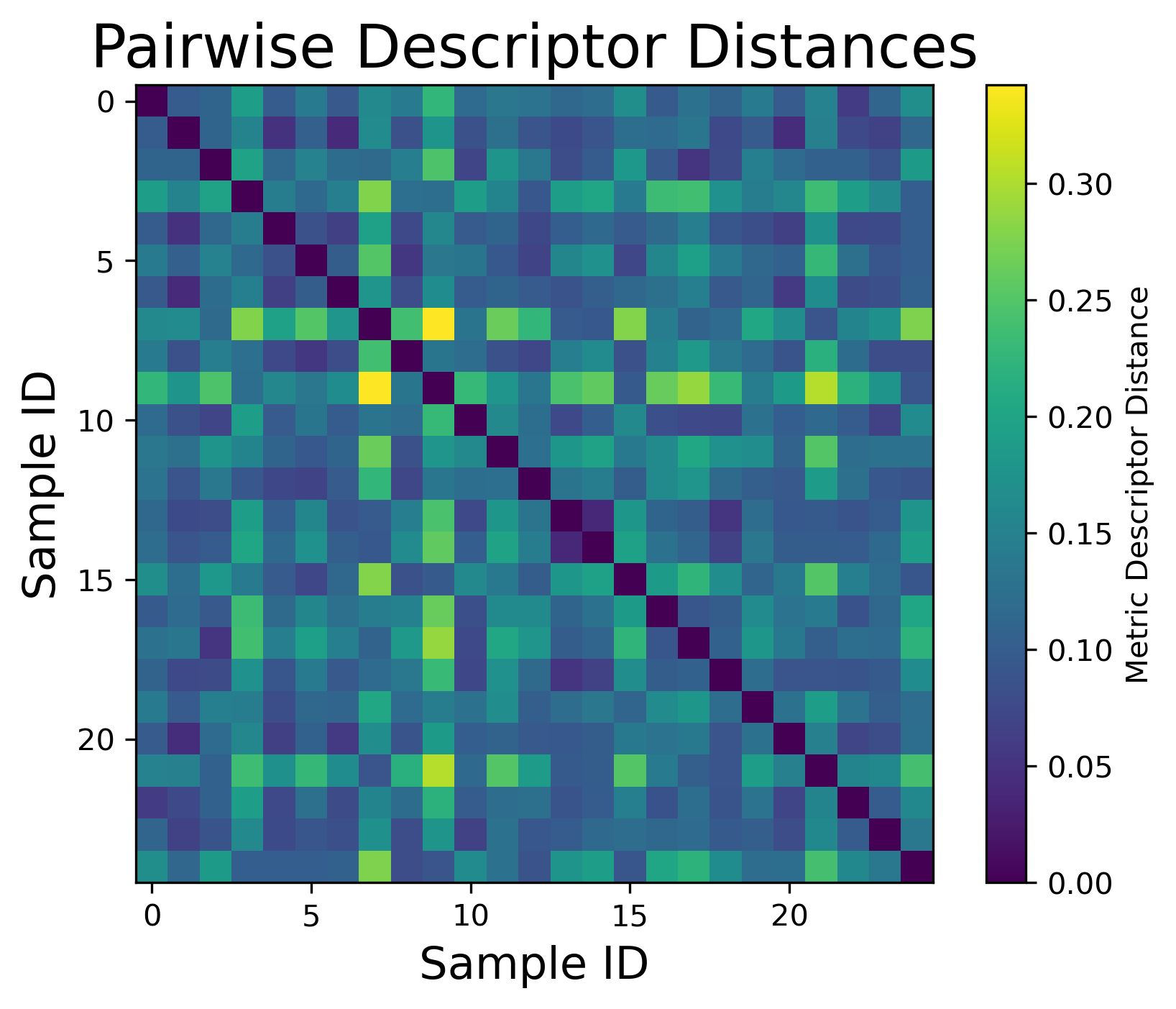} \caption{} \label{fig:mds_pairwise} \end{subfigure}  \hfill \caption{(a) MDS projection of $m$-instances within a $M$-state and corresponding (b) pairwise metric descriptor distances array used to determine MDS positions.} \label{fig:mds} \end{figure}

Multidimensional scaling (MDS) is a non-linear dimensionality reduction technique that can be used to visualize the similarities between $m$-instances sampled from a given $M$-state by determining the Cartesian coordinates associated with each data point using the pairwise metric distance values between all samples within a population \cite{carroll1998multidimensional}. Figure \ref{fig:mds_kde} shows how MDS can be used to determine spatial locations in the embedded domain from a set of $m$-instances and how kernel density estimation (KDE) can be used to visualize the distribution within one $M$-state. 
 Figure \ref{fig:mds_pairwise} shows the pairwise MDD values between each of the $m$-instances in the set used to determine the MDS projection. 

\subsection{Moving from Local to Global Manifold Relationships}

\begin{figure}[!htb] \begin{centering} \includegraphics[width=0.75\linewidth]{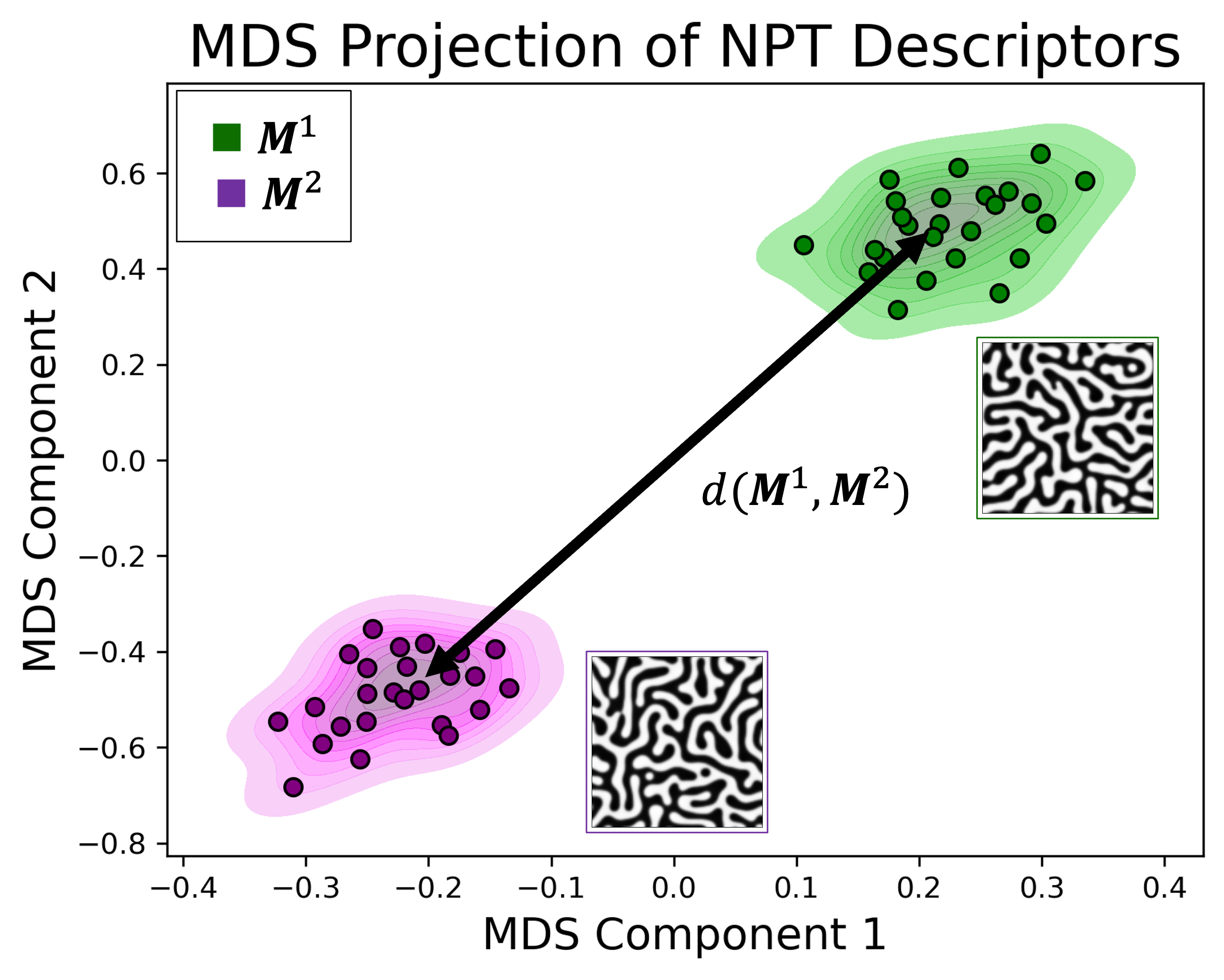} \caption{MDS projection of two neighboring $M$-states and their distributions of $m$-instances.} \label{fig:mds_M1M2_kde} \end{centering} \end{figure}

When expanding analysis across the manifold, interest shifts from comparing individual $m$-instances to comparing $M$-states. While the underlying $M^\theta$ stochastic process is unknowable, it can be estimated as $\hat{M}^\theta$ from the available information -- sets of $m$-instances. While there are many potential choices for approximating $M$-states and establishing distance metrics between distributions, such as the Earth-Mover distance used by Miley et al. \cite{miley_microstructure_2025}, for simplicities sake, this work will approximate the $M$-state information as the average descriptor for a set of $m$-instances and maintain the p-Minkowski-based MDD to calculate distances between material states. Equation \ref{eq:Mhat} shows this method for estimating a descriptor of the $M$-state, $\hat{M}^\theta$:

\begin{equation} \label{eq:Mhat} \hat{M}^\theta \approx \bar{D}^\theta = \frac{1}{n}\sum_{j=1}^{n} D(m_j^\theta) \end{equation}

where $\bar{D}^{\theta}$ is the average descriptor vector calculated on a set of $m$-instances. Figure \ref{fig:mds_M1M2_kde} shows an example of two neighboring $M$-states including their respective distributions of $m$-instances in comparison to one another.

% Establishing distance metrics between $m$-instances and $M$-states is necessary for understanding microstructure as a stochastic process \cite{desaiCOMODOConfigurableMorphology2024}. While distances between $M$-states can be calculated based on their $\theta$ parameters and may offer some semblance of expected similarity between two $M$-states, those distances may not be informative of how microstructures exist on the manifold. There is no sense of variation within one $M$-state, so it is not possible to assert questions like ``what is the structurally weakest state that exists within this population?'' Additionally, there may be areas of the $\mathcal{M}$-manifold where a small change in the $\theta$-parameter leads to a relatively larger change in the $M$-state, defining regions of stability or instability. However, that information can be captured through measurements on the resultant $M$-states.

\begin{figure}[!htb] \begin{centering} \includegraphics[width=\linewidth]{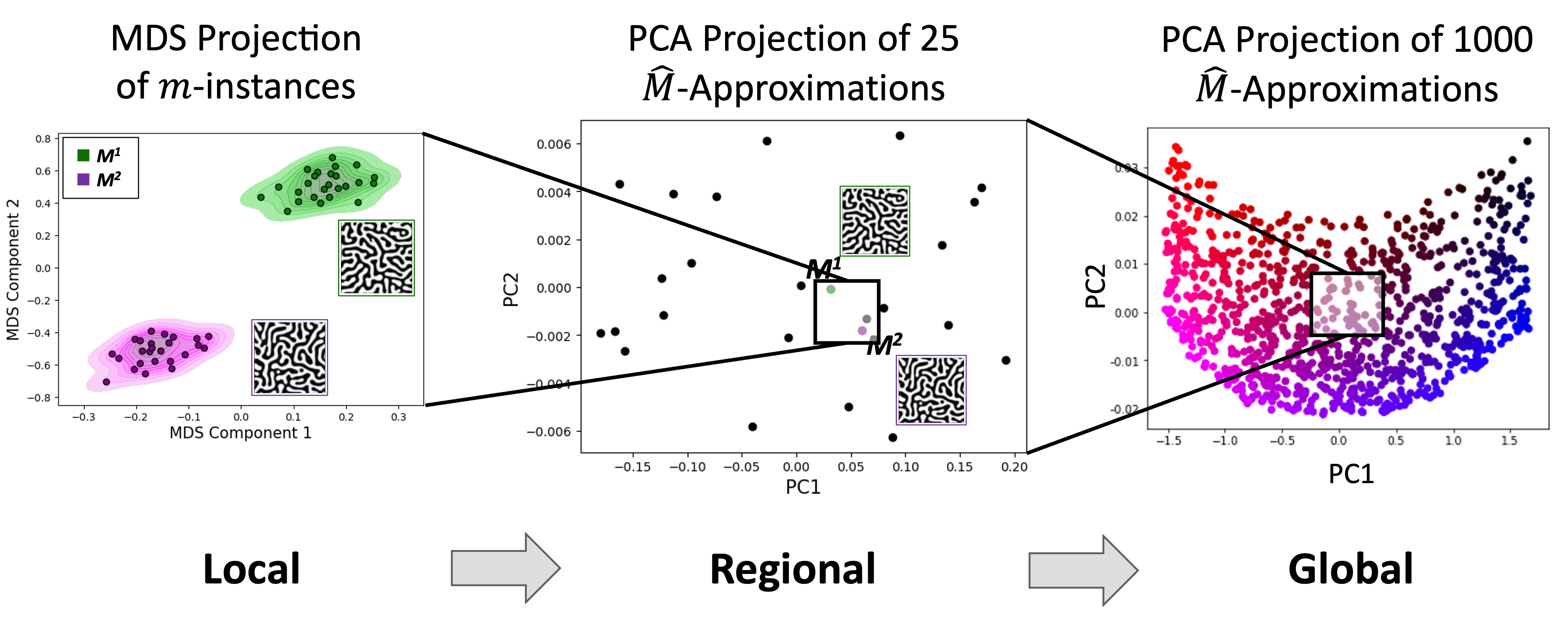} \caption{Expanding manifold analysis from local to regional to global relationships.} \label{fig:local2global} \end{centering} \end{figure}

In understanding the $\mathcal{M}$-manifold as the span of $M$-states within a processing domain, manifold assessment can be done on local comparisons between $m$-instances, regional comparisons between neighborhoods of similar $M$-states, and global assessments on all $M$-states in the domain. Figure \ref{fig:local2global} shows manifold projections of these three windows, using the MDS projection for local assessment, and principal component analysis (PCA) to visualize the relationships between $\hat{M}$-approximations for regional and global families of data. PCA and the manifold visualization techniques are described in detail in Section \ref{sec:mani_vis}.

\subsection{Properties of the Material Manifold}

As presented in the introduction, dimensionality and invertibility are used to assess the quality of the low-dimensional manifold mapping between processing and structure domains. These notions act complementarily to each other to establish the network of $M$-states across the $\mathcal{M}$-manifold. \vspace{\baselineskip}

The \textbf{dimensionality}, or more specifically, the intrinsic dimensionality of our material system, is of great interest in determining the minimum number of parameters needed to define the manifold, thereby reducing the complexity in future analysis and application tasks. The intrinsic dimensionality is the number of controlling dimensions required to describe a system space. Dimensionality assessment is performed to determine if the microstructure representations are capturing information on the underlying materials processes rather than the aleatoric noise that occurs due to the stochastic nature of sampling multiple microstructure realizations at each $M$-state. This test is completed by verifying that the mapping from the processing domain, $\Theta$, to the structure domain, $M$, is dimensionality-preserving - or that $\Theta$ and $M$ both exist in $\mathbb{R}^n$.

The \textbf{invertibility} assessment is performed to validate that the chosen microstructure descriptors are capturing salient information on material states and microstructural features present within the $m$-instances in such a way that their corresponding $\theta$-parameters can be recovered.

These properties are inherently linked, and by assessing both of these properties we can work to establish a meaningful, low-dimensional representation of the $\mathcal{M}$-manifold that can be integrated with the manifold hypothesis.

%%%%%%%%%%%%%%%%%%%%%%%%%%%%%%%%%%%%%%%%%%%%%%%%%%%%%%%%%%%%

\section{Methods}
\subsection{Spinodal Decomposition Phase Field Simulations} \label{sec:pf_system}

The material domain explored in this work is a representation of the spinodal decomposition phase separation mechanism in two-phase material systems. This is a thermodynamic process by which complex microstructures are generated when a liquid melt having a composition inside the spinodal region is quenched to form a solid, leading to spontaneous phase separation. This quenching process and resulting microstructure is accurately represented by numerically solving the Cahn-Hilliard equation \cite{chenContinuumFieldApproach1996}, \cite{blomkerSpinodalDecompositionforCahn2001} using Equation \ref{eq:cahn}:

% \begin{equation}  \label{eq:cahn}  \begin{array}{ll}  \frac{\partial c}{\partial t} = -\Delta(\kappa^2 \Delta c + f(c)) + \sigma * \xi & \text{in } G\\  \frac{\partial c}{\partial \nu} = \frac{\partial \Delta c}{\partial \nu} = 0 & \text{on } \partial G  \end{array}  \end{equation}

\begin{equation} \label{eq:cahn} \frac{\partial c}{\partial t} = D\nabla^2(c^3 - c -\kappa\nabla^2c) \end{equation}

where $c$ is the local relative concentration of the binary system components, $D$ is the diffusivity, and $\kappa$ is the atomic scale interaction length. In this case, $\kappa$ acts as a proxy for interfacial energy, wherein larger values of $\kappa$ correspond to a greater energy penalty for having a boundary between phases, leading to fewer, larger particles. The initial stages of separation are highly dependent on the volume fraction, $\eta$, which controls the mean value of Gaussian random noise field and thus the local concentration, $c$, across the spatial realization of microstructure.

The phase-field simulations were run using a C++ finite difference solver under the conditions of a symmetric double-well free energy potential assuming fully-coherent interfaces between phases. Diffusivity was fixed for all simulations, and simulations were run for the same time-scale for $t \in [0,T]$. Final time $T$ was predetermined as the length of time needed for microstructure state to remain static over a given $\Delta t$. Microstructure simulations were generated on a 2D square grid with size $256 \times 256$ pixels for dimensionless spatial and temporal parameters with periodic boundary conditions over all boundaries of the domain. Images are grayscale with uint8 bit depth. The simulations were controlled by three parameters: $\eta$, $\kappa$, and $\xi$. Volume fraction, $\eta$, and interfacial energy, $\kappa$, serve as a proxy for material composition and processing conditions to define the materials process. Parameters uniformly sampled for $\eta$ in the range [-0.25, 0.25] and $\kappa$ in the range [0.25,0.75]. Initial conditions were defined as a Gaussian noise filed in the $256 \times 256$ image domain, seeded by $\xi$, a random integer. $\xi$ can be interpreted as the stochastic effects of thermal fluctuations in the spatial distribution of phase concentration.

\begin{figure}[ht!] \begin{centering} \includegraphics[width=0.75\linewidth]{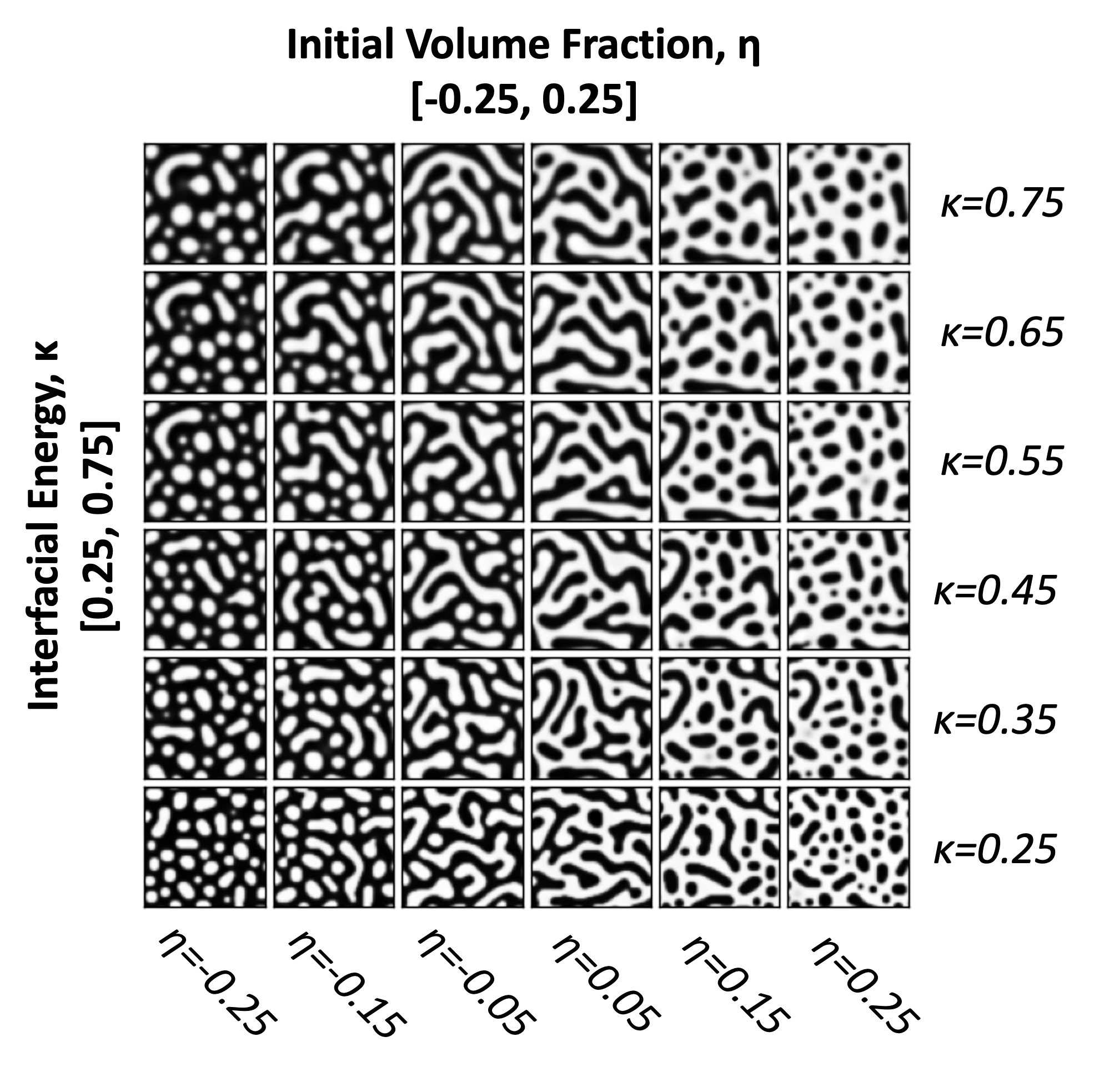} \caption{Phase field material manifold from input conditions $\eta$ and $\kappa$.} \label{fig:param_space} \end{centering} \end{figure}

Images generated from these simulations across the entire span of $\Theta$ allows for a wide range of microstructural features to be represented. The span of input parameters and resulting family of $M$-states builds the material domain on which to explore how changing the processing parameters impacts the resulting microstructure. Figure \ref{fig:param_space} shows example microstructures at corresponding $\theta$-parameter pairs across the processing domain, where each individual microstructure was generated with the same random seed, $\xi$, where the spatial location of features is similar from one microstructure to the next, giving visual parity for easier interpretation of differences in microstructures across the domain.

\subsection{Manifold Sampling and Stochastic Approximation of M-State Information} \label{sec:datasetSampling}

In exploring the $\mathcal{M}$-manifold, a dataset of 25,000 microstructure images was generated by uniformly sampling the $\Theta$-processing space according to a Poisson point process for a set of 1000 $M^\theta$-states. For each $M^\theta$-state, 25 $m$-instances were generated by seeding the initial concentration $c$, from a Gaussian noise-field, $\xi$, having a mean value of $\eta$ and numerically solving the Cahn-Hilliard equation for that set of $(\eta,\kappa,\xi)$ initial conditions.

The choice of 1000 $M$-states was made \textit{a priori}, as this is a sufficient number of points needed for a reliable dimensionality estimation of the $\mathcal{M}$-manifold. In order to balance dataset size and simulation time against the convergence of the microstructure descriptors, 25 $m$-instances were generated for each $M$-state, justified \textit{a posteriori} by the error bands on the estimate of the average microstructure descriptor.

\begin{figure}[!ht] \begin{centering} \includegraphics[width=0.8\linewidth]{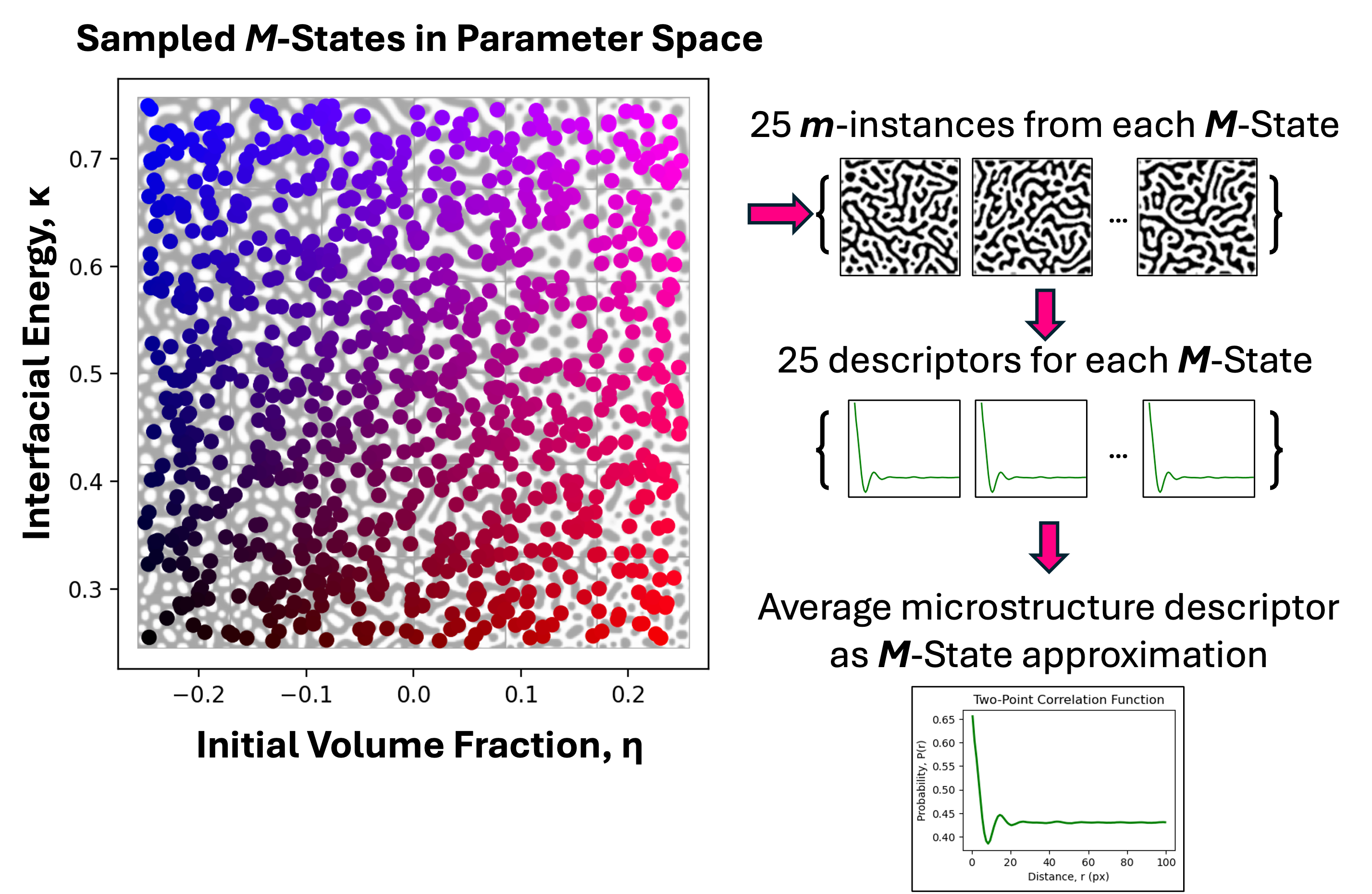} \caption{Processing parameter-material manifold with sampled $M$-states colored by red/blue parameter pair.} \label{fig:PFmani_sampling} \end{centering} \end{figure}

Using Equation \ref{eq:Mhat}, $\hat{M}^\theta$ approximations were calculated as the average microstructure descriptors, $\bar{D}^{\theta}$. Individual descriptor vectors were calculated for each of the 25,000 $m$-instances, and approximations of the $M$-state descriptors were calculated by averaging each of the 25 $m$-instances in each of the 1000 $M$-states. Figure \ref{fig:PFmani_sampling} shows the sampled states across the processing domain, with each point representing $M$-state defined by a pair of processing parameters. Each point is colored according to its $\theta$-parameters, with the amount of red corresponding to the $\eta$ parameter and the amount of blue corresponding to the $\kappa$ parameter. This color scheme is used throughout this work as a tool for visual parity in understanding the different data projections of the processing, structure, and material manifolds.

Revisiting the processing-structure feedback loop from Figure \ref{fig:tikz_simple}, while physically the forward process goes directly from the processing domain to the structure domain, the material state is not known until descriptors are calculated on $m$-instances from the generating process (phase-field), and the $M$-state descriptor is approximated. The inverse process, $g$, maps back from $\hat{M}$ to $\hat{\Theta}$, or the recovered parameters. This revised diagram is shown in Figure \ref{fig:tikz_complex}.

\begin{center}
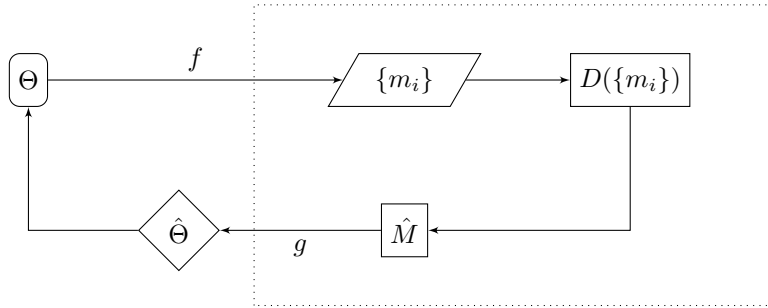
 \begin{tikzpicture}[auto,node distance=2cm,>=latex']

\node [terminator] (param) at (-7, 0) {$\Theta$};

\node [data] (micro) at (-2, 0) {$\{m_i\}$}; \node [process] (desc) at (1, 0){$D(\{m_i\})$}; \node [process] (bigM) at (-2, -2) {$\hat{M}$};

\node [decision] (update) at (-5, -2) {$\hat{\Theta}$}; \node [point, below of=param] (belowParam) {}; \node [point, below of=desc] (belowDesc) {};

\draw [->] (param) -- node {$f$} (micro); \draw [->] (micro) -- (desc); \draw [->] (desc) -- (belowDesc) -- (bigM); \draw [->] (bigM) -- node {$g$} (update); \draw[->] (update.west) -- (belowParam) -- (param.south);

\draw[dotted] (-4,-3) rectangle (3,1);

\end{tikzpicture} \captionof{figure}{Updated invertible mapping flowchart through stochastic approximation of $M$-state information.} \label{fig:tikz_complex} \end{center}

\subsection{Microstructure Descriptors} \label{sec:microDescriptor}

Using the dataset described in Section \ref{sec:datasetSampling}, three different reduced-order microstructure descriptors were calculated to compare the quality of descriptors in representing individual microstructures, as well as in constructing the material manifold itself. The descriptors chosen were persistent homology (PH), two-point correlation function (NPT), and average chord length (ACL). Descriptor calculations were done on each image, reducing the dimensionality of the $m$-instances from the ambient, image space, $m \in \mathbb{R}^{256 \times 256}$, to that of the descriptor space: $D_{PH} \in \mathbb{R}^{2 \times 100}$, $D_{NPT} \in \mathbb{R}^{1 \times 100}$, and $D_{ACL} \in \mathbb{R}^{2}$ respectively.

\noindent\textbf{Persistent Homology:}

Persistent homology is a topological data analysis technique that is able to summarize the connectivity of microstructures over a range of length-scales \cite{zomorodianComputingPersistentHomology2005}. Persistence silhouettes are one method in which persistent homology data can be summarized \cite{chazal2014stochastic} and were chosen for this work because of the vector-based representation of persistent homology data. This allowed for simple integration with the computational and statistical analysis methods. The Geometry Understanding in Higher Dimensions (GUDHI) \cite{gudhi:urm} and giotto-tda \cite{tauzin2021giotto} python packages were used to perform PH calculations.

\begin{figure}[ht] \begin{centering} \includegraphics[width=0.8\linewidth]{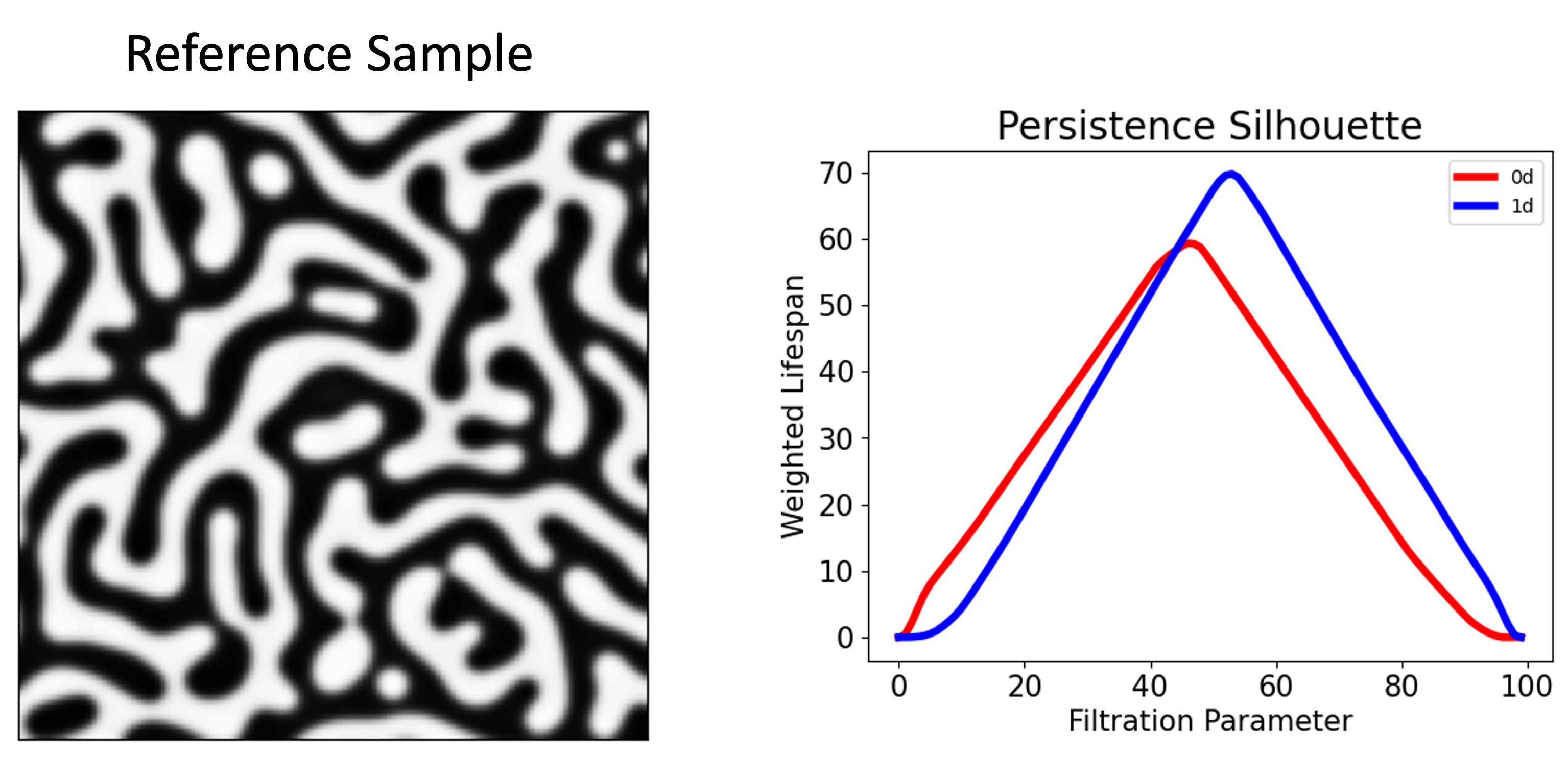} \caption{Reference microstructure and corresponding persistence silhouette.} \label{fig:ph_desc} \end{centering} \end{figure}

Persistence calculations were done on each image in ambient space $m \in \mathbb{R}^{256 \times 256}$, reducing the dimensionality to that of the silhouette, $D_{PH} \in \mathbb{R}^{2 \times 100}$. First, the persistence diagram for each image was calculated with giotto-tda's `CubicalPersistence' function. The diagram was calculated for a periodic cubical complex on both the x and y axes, given that the phase field simulations themselves were periodic. From there, the silhouettes were calculated using the `Silhouette' function, with a power weighting of 1.0 (unweighted), and with a bin size of 100 - returning two $1 \times 100$ silhouette curves, one for each homology dimension. From there, the $M$-state average silhouettes were calculated for the 0D and 1D curves, resulting in the final representation of $D_{PH} \in \mathbb{R}^{2 \times 100}$. Figure \ref{fig:ph_desc} shows an example of a microstructure and its corresponding persistence silhouette, with one silhouette curve for the 0D topological features (components/black phase) and a second curve for the 1D topological features (holes/white phase).

\noindent\textbf{Two-Point Statistics:}

N-point correlation statistics are a commonly used method for microstructure informatics \cite{niezgodaOptimizedStructureBased2010}. The two-point correlation function describes the probability of finding vectors of random positions, directions, and magnitudes that have heads/tails in the same phase. This information is able to describe microstructural information like precipitate width and spacing, as well as capture averaged information like volume fraction.

\begin{figure}[ht] \begin{centering} \includegraphics[width=0.8\linewidth]{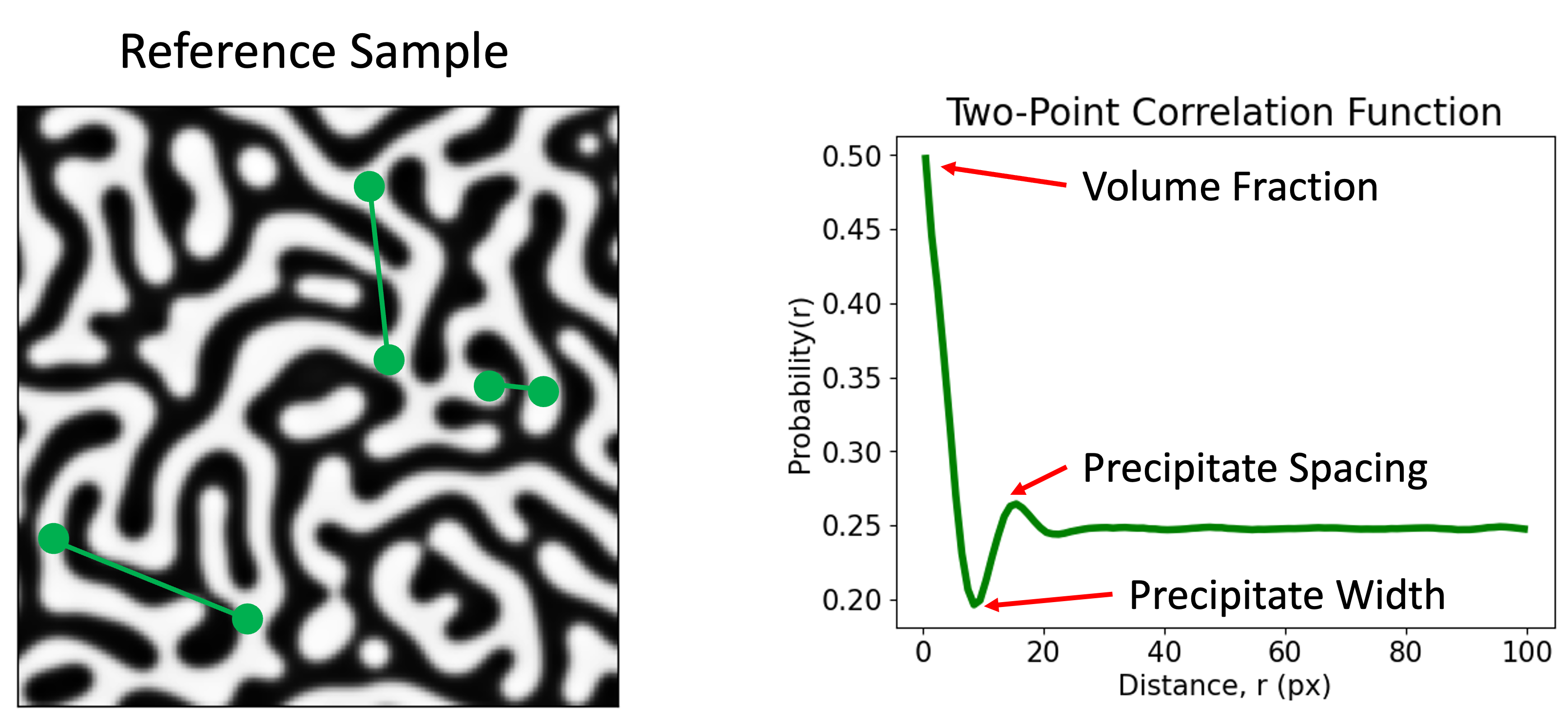} \caption{Reference microstructure and corresponding 2-point autocorrelation function.} \label{fig:npt_desc} \end{centering} \end{figure}

Similarly to the persistence calculations, the 2-point correlation function was used to find a reduced-order microstructure descriptor, going from the ambient dimensionality of $m \in \mathbb{R}^{256 \times 256}$ down to the radially-integrated autocorrelation vector, $D_{NPT} \in \mathbb{R}^{1 \times 100}$. PoreSpy \cite{Gostick2019} was used to calculate the descriptor vectors using the `two\_point\_correlation' function. The images were first binarized before calculating the autocorrelation vector with periodic boundaries on the x and y dimensions. The dataset used in this work contains purely-isotropic microstructures, meaning the vector form of the NPT function can be used as opposed to the full-field image. Figure \ref{fig:npt_desc} shows an example of a microstructure and its corresponding two-point autocorrelation function.

\noindent\textbf{Average Chord Length:}

The average chord length was included as an alternative microstructure descriptor to test the manifold recovery qualities of a greatly--reduced--order descriptor. Due to the nature of its calculation, the ACL is inherently a local-state descriptor and cannot capture the effects of long-range order in a microstructure. It was initially chosen as a simplification of the chord length density (CLD) function and was expected to fail at manifold recovery due to the extreme reduction in dimensionality.

In contrast to NPT, CLD calculates the probability of a chord with length $r$ contained wholly within a single phase \cite{torquatoSTATISTICALDESCRIPTIONMICROSTRUCTURES2002}. While the full CLD function is useful for assessing connectivity properties like percolation, reducing that second-order representation from a distribution of measurements down to a first-order average with the ACL limits the description of microstructural features down to, on average, ``how wide are precipitates'' and ``how much spacing is there between precipitates''.

\begin{figure}[ht] \begin{centering} \includegraphics[width=0.8\linewidth]{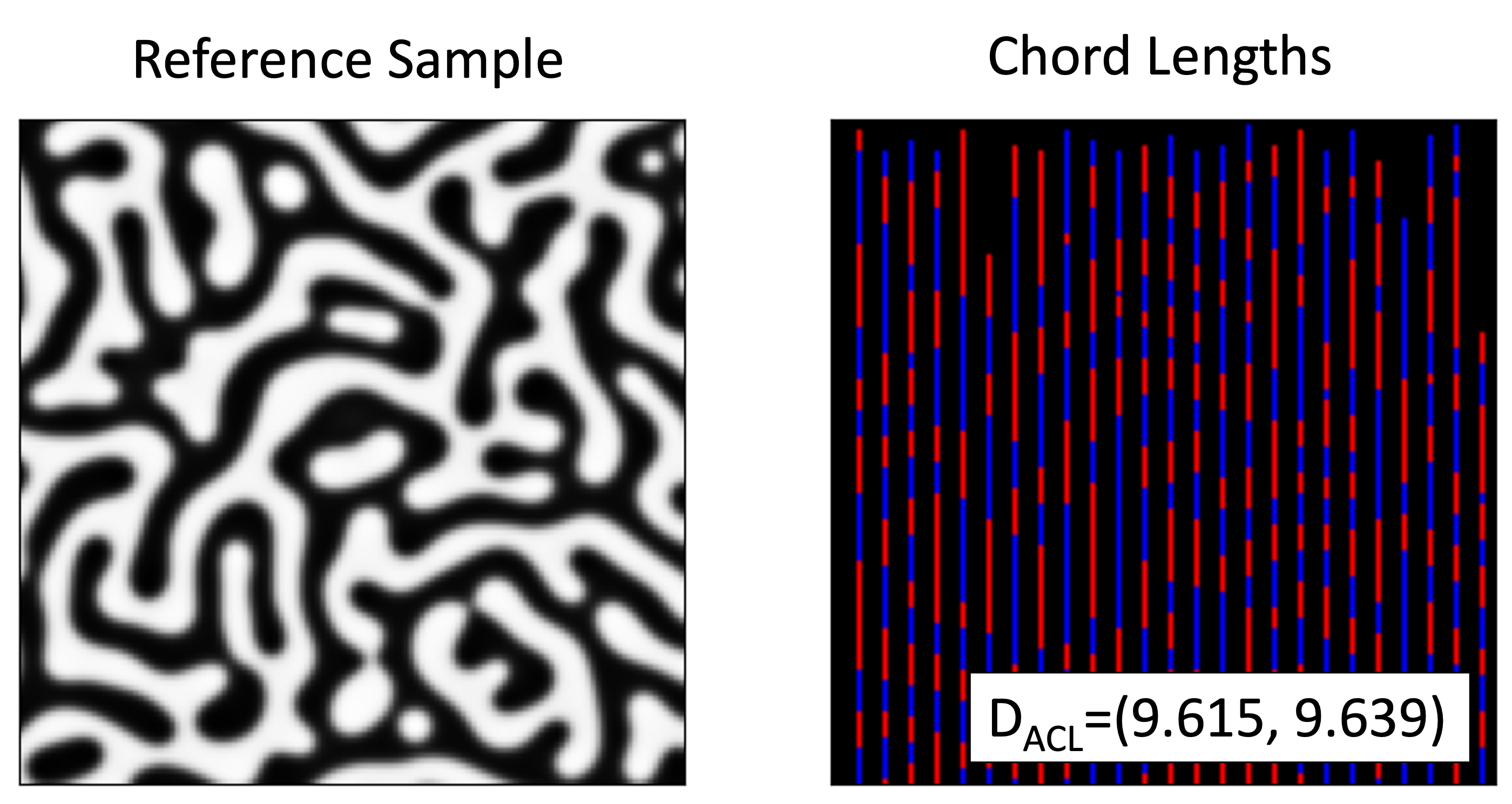} \caption{Reference microstructure and corresponding chord lengths for black phase (shown in blue) and white phase (shown in red).} \label{fig:acl_desc} \end{centering} \end{figure}

Two different representations of the ACL descriptor were compared: the ACL of a single phase, arbitrarily chosen as the white phase, and the ACL of both phases. These two descriptors are notated as ``ACL-1'' and ``ACL-2''. PoreSpy was used for the ACL calcuations, calculating a reduced-order descriptor in $D_{ACL-1} \in \mathbb{R}^{1}$ and $D_{ACL-2} \in \mathbb{R}^{2}$. The `apply\_chords' function was used to calculate the length of white-phase vectors in one orthogonal direction. From there, the `count\_chords' function was used to count the length of chords applied in the previous function and averaged to return the average chord length. This process was repeated on the image inverse to determine the average chord length of the black phase. Again, because the dataset is isotropic, the descriptor was only calculated for a single orthogonal direction. Figure \ref{fig:acl_desc} shows an example of a microstructure and its corresponding average chord lengths for the black and white phases, with chords drawn over the reference sample shown in blue and red for the black and white phases respectively.

%%%%%%%%%%%%%%%%%%%%%%%%%%%%%%%%%%%%%%%%%%%%%%%%%%%%%%%%%%%% %%%%%%%%%%%%%%%%%%%%%%%%%%%%%%%%%%%%%%%%%%%%%%%%%%%%%%%%%%%% 
\section{Results}
\subsection{Dimensionality Estimation} \label{sec:dimEst}

The estimation of system dimensionality through a nearest-neighbor approach is used as one method for assessing the quality of the microstructure descriptors. In this work, dimensionality estimation was used as verification method to determine the quality of the chosen descriptors, but could alternatively be applied to estimate the dimensionality of an unknown system in an effort to determine the number of controlling parameters. Developed by \cite{levina2004maximum} and applied for material manifold exploration by \cite{sundararaghavanMicrostructuralStateVariables2023}, the Maximum Likelihood Estimator (MLE)-based nearest-neighbor dimensionality estimation technique recovers the intrinsic dimensionality of a system. This method relies on having a true distance metric between $M$-states to construct the nearest-neighbor network between microstructure states. For the spinodal system used in this work, it is known that its intrinsic dimensionality is two, as the entire process can be described by changing the two input $\theta$-parameters, $\eta$ and $\kappa$. The third parameter, $\xi$, has a major impact on the resulting micrographs, but changes only the aleatoric dimensinoality, not the intrinsic.

For the chosen microstructure descriptors, $1000\times1000$ matrices of pairwise MDD values between each of the $\hat{M}$-approximations were calculated and used in the estimation of manifold dimensionality. Three different assessments were completed: a comparison of 1000 little-$m$ microstructures having the same initial seed, 1000 little-$m$ microstructures having random initial seeds, and a comparison of the $\hat{M}$-approximations of the full dataset. The two little-$m$ comparisons were calculated to show the inability of the direct image distance to capture meaningful microstructure information when stochastic variation is introduced to the system. For the same seed dataset, little-$m$ samples with similar $\theta$-parameters have similar spatial information. In this case, the direct image distance is able to construct a correct nearest-neighbor network, resulting in an accurate dimensionality estimation of 2.15, as shown in Table \ref{tab:dimest_sslm}. This descriptor distance falls apart when the images are generated using a random seed, changing the initial state of the simulation and the resulting microstructures have no shared spatial information with results shown in Table \ref{tab:dimest_rslm}. In this case, the dimensionality estimation returns 119.86, lower than the ambient dimensionality of $\mathbb{R}^{65536}$ for $256\times256$ images, but an intractable representation for manifold construction and application due to the failure to capture aleatoric noise.

\begin{table}[!htb] \begin{minipage}{.5\linewidth} \centering \caption{Same Seed, Little $m$\\MLE Dimensionality Estimation} \label{tab:dimest_sslm} \begin{tabular}{lccl} \toprule \textbf{Same Seed, Little $m$} \\ & $\mu$ & $\sigma$ \\ \cmidrule{2-3} Direct Image Distance & 2.15 & 0.55 \\ PH Distance & 2.80 & 1.16 \\ NPT Distance & 2.07 & 0.65 \\ ACL-2 Distance & 1.92 & 0.67 \\ ACL-1 Distance & 1.00 & 0.31 \\ \bottomrule \end{tabular} \end{minipage} \begin{minipage}{.5\linewidth} \centering \caption{Random Seed, Little $m$\\MLE Dimensionality Estimation} \label{tab:dimest_rslm} \begin{tabular}{lccl}\toprule \textbf{Random Seed, Little $m$} \\ & $\mu$ & $\sigma$ \\ \cmidrule{2-3} Direct Image Distance & 119.86 & 41.00 \\ PH Distance & 2.74 & 1.00 \\ NPT Distance & 2.22 & 0.67 \\ ACL-2 Distance & 1.98 & 0.61 \\ ACL-1 Distance & 1.00 & 0.33 \\ \bottomrule \end{tabular} \end{minipage} \end{table}

\begin{figure}[!ht] \centering \begin{subfigure}[c]{0.45\textwidth} \includegraphics[width=\textwidth]{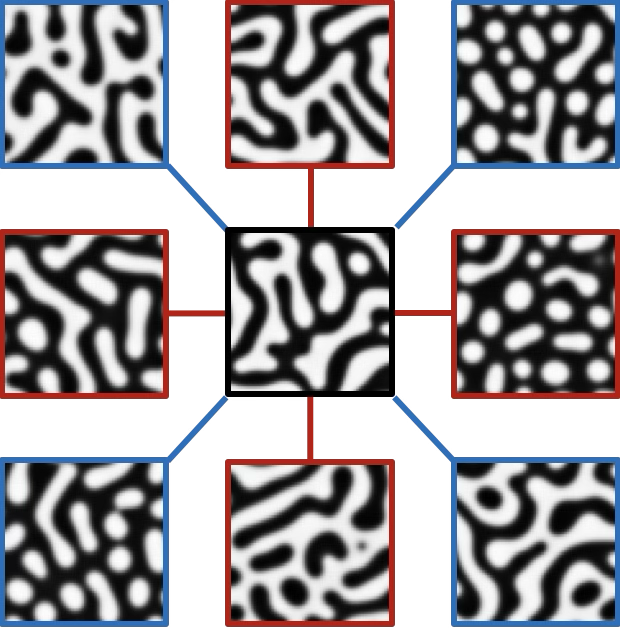} \caption{} \label{fig:nn_did} \end{subfigure} \hfill \begin{subfigure}[c]{0.45\textwidth} \includegraphics[width=\textwidth]{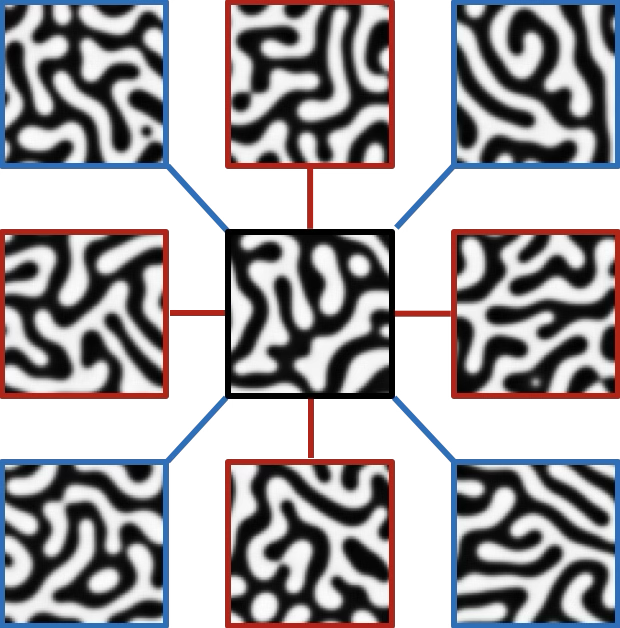} \caption{} \label{fig:nn_npt} \end{subfigure}  \hfill \caption{First (red) and second (blue) nearest neighbor shells as determined using the MDD calculation for (a) direct-image and (b) $D_{NPT}$ distances.} \label{fig:nn_shells} \end{figure}

Figure \ref{fig:nn_shells} shows examples of nearest-neighbor relationships of $m$-instances determined for the case of micrographs with random seeds, as used for Table \ref{tab:dimest_rslm}, for the (a) direct-image and (b) $D_{NPT}$ distances. These images show the failure of the direct image distance to establish meaningful relationships between $m$-instances, contributing to the high estimated dimensionality and inability to uncover a meaningful manifold.

When utilizing all 25 $m$-instances to calculate the $\hat{M}$-approximation, the results show that the three main descriptors ($D_{PH}$, $D_{NPT}$, and $D_{ACL-2}$) are able to successfully recover the true intrinsic dimensionality of the system, shown in Table \ref{tab:dimest_bigm}, successfully capturing the stochasticity of the system and recovering a manifold having the minimum amount of parameters necessary. To visualize this, Section \ref{sec:mani_vis} plots 2D representations of the $\hat{M}$-approximation domain. The $D_{ACL-1}$ descriptor recovers a dimensionality of 1, but that is to be expected as the $D_{ACL-1}$ descriptor itself is only a single value and cannot define a domain in higher dimensions.

\begin{table}[ht] \begin{center} \caption{Big $M$ Nearest-Neighbor MLE Dimensionality Estimation} \label{tab:dimest_bigm} \begin{tabular}{lccl} \toprule \textbf{Random Seed, Big $M$} \\ & $\mu$ & $\sigma$ \\ \cmidrule{2-3} PH Distance & 2.08 & 0.86 \\ NPT Distance & 1.83 & 0.57 \\ ACL-2 Distance & 1.94 & 0.60 \\ ACL-1 Distance & 1.00 & 0.32 \\ \bottomrule \end{tabular} \end{center} \end{table}

\subsection{Invertibility Assessment: Parameter Recovery using KNN Regression} \label{sec:invert}

In order to verify that the chosen descriptors are capturing and encoding representative $M$-state information, two regression models were trained to recover the input parameters to the phase field model used to generate the microstructures themselves. Support vector regression (SVR) and K-Neighbors Regression (KNR) are supervised regression models capable of predicting target values ($\theta$-parameters for phase field model) from the predictor variables (microstructure $D$-descriptor vectors). For this analysis, the dataset of 1000 $M$-states was separated into training/testing subsets using an 80/20 split. Scikit-Learn's `LinearSVR' and `KNeighborsRegressor' models were used with default model parameters \cite{scikit-learn}. For the `LinearSVR' model, $D$-descriptor vectors were passed directly as the predictor variables. For the `KNeighborsRegressor' model, $1000\times1000$ pairwise MDD matrices for each of the descriptor methods were passed in using the function's `precomputed' flag.

Tables \ref{tab:lsvr_r2}, \ref{tab:lsvr_rmse}, \ref{tab:KNR_r2}, and \ref{tab:KNR_rmse} show regression model results in terms of the models' coefficients of determination ($R^2$) and root mean squared error (RMSE) values. The $R^2$ value shows how well a model fits the dataset by reporting the proportion of the variation in the dependent variable that is predictable from the independent variable. The RMSE reports on average how far apart the predicted values were from the true values and can be directly interpreted in terms of $\Delta \theta$, or the average difference in recovered parameters.

\begin{table}[!htb] \begin{minipage}{.5\linewidth} \centering \caption{LinearSVR - $R^2$} \label{tab:lsvr_r2} \begin{tabular}{lccc} \toprule & & \multicolumn{2}{c}{Parameters} \\ \cmidrule{3-4} & Combined & $\eta$ & $\kappa$ \\ \midrule PH & 0.9512 & 0.9361 & 0.9663 \\ NPT & 0.9919 & 0.9992 & 0.9846 \\ ACL-2 & 0.9786 & 0.9924 & 0.9647 \\ ACL-1 & 0.3178 & 0.6237 & 0.0120 \\ \bottomrule \end{tabular} \end{minipage}\begin{minipage}{.5\linewidth} \centering \caption{LinearSVR - RMSE} \label{tab:lsvr_rmse} \begin{tabular}{lccc} \toprule & & \multicolumn{2}{c}{Parameters} \\ \cmidrule{3-4} & Combined & $\eta$ & $\kappa$ \\ \midrule PH & 0.0313 & 0.0351 & 0.0269 \\ NPT & 0.0132 & 0.0040 & 0.0182 \\ ACL-2 & 0.0213 & 0.0121 & 0.0276 \\ ACL-1 & 0.1195 & 0.0853 & 0.1458 \\ \bottomrule \end{tabular} \end{minipage} \end{table}

\begin{table}[!htb]  \begin{minipage}{.5\linewidth} \centering \caption{KNR - $R^2$} \label{tab:KNR_r2}   \begin{tabular}{lccc} \toprule & & \multicolumn{2}{c}{Parameters} \\ \cmidrule{3-4} & Combined & $\eta$ & $\kappa$ \\ \midrule PH & 0.9811 & 0.9893 & 0.9729\\ NPT & 0.8729 & 0.9998 & 0.7459 \\ ACL-2 & 0.9933 & 0.9990 & 0.9875 \\ ACL-1 & 0.4385 & 0.8451 & 0.0318 \\ \bottomrule \end{tabular} \end{minipage} \begin{minipage}{.5\linewidth} \centering \caption{KNR - RMSE} \label{tab:KNR_rmse} \begin{tabular}{lccc} \toprule & & \multicolumn{2}{c}{Parameters} \\ \cmidrule{3-4} & Combined & $\eta$ & $\kappa$ \\ \midrule PH & 0.0199 & 0.0144 & 0.0241 \\ NPT & 0.0523 & 0.0018 & 0.0740 \\ ACL-2 & 0.0120 & 0.0044 & 0.0164 \\ ACL-1 & 0.1092 & 0.0547 & 0.1444 \\ \bottomrule \end{tabular} \end{minipage} \end{table}

When comparing the $R^2$ values of the two models, it shows the $D_{PH}$, $D_{NPT}$, and $D_{ACL-2}$ descriptors all succeed in their abilities to recover the $\theta$-parameters. The LinearSVR results, shown in Table \ref{tab:lsvr_r2} show the $D_{NPT}$ descriptor as the highest-performing model, with superb recovery of the $\eta$ parameter, and very high recovery of the $\kappa$ parameter. The $\eta$ parameter's $R^2$ value of $0.9992$ is not unexpected for the $D_{NPT}$ descriptor, as the volume fraction of the microstructure images are explicitly encoded in the descriptor vector. However, when comparing the $D_{NPT}$'s LinearSVR results to the KNR model, the $D_{NPT}$ descriptor becomes the lowest-performing descriptor of the three ($D_{ACL-1}$ excluded). One potential reason for this discrepancy in result could be due to the nature of the $D_{NPT}$ distance metrics in constructing a nearest-neighbor network to predict parameters. The KNR model has a similar $R^2$ for the $\eta$ parameter, but drastically decreases for the $\kappa$ parameter. We hypothesize that this results from the relative sensitivity of the $D_{NPT}$ descriptor regarding $\eta$, where differences in $\eta$ lead to distances orders of magnitude greater than differences in $\kappa$. Thus, when the KNR model is predicting parameters based off of the nearest neighbors regarding MDD distance, it can much more readily distinguish between $M$-states with different $\eta$ values.

The invertibility assessment was performed with a parameter recovery goal of RMSE$\leq$0.05, as this is a $\Delta\theta$ threshold at which practitioners can begin to visually differentiate between $M$-states. Aside from the $D_{NPT}$'s lack of sensitivity in the $\kappa$ parameter for the KNR model, the three main descriptors ($D_{PH}$, $D_{NPT}$, $D_{ACL-2}$) all meet this minimum RMSE threshold.

Looking at the performance of the $D_{ACL-2}$ and $D_{ACL-1}$ descriptors, we expected to see the $D_{ACL-1}$ descriptor fail to recover the input parameters due to both the limited information available to the model in the $D_{ACL-1}$ descriptor, as well as the ill-posed nature of attempting to recover data in $\mathbb{R}^2$ from $\mathbb{R}^1$. Along with the $D_{ACL-2}$'s ability to recover the system dimensionality, it was quite surprising to see just how well the $D_{ACL-2}$ descriptor performed at the invertibility assessment, on average outperforming both the $D_{PH}$ and $D_{NPT}$ descriptors.

While there is certainly more that can be done to improve these regression model results in terms of model selection and hyperparameter tuning, the inclusion of two different models in this section is to illustrate how changing the inverse model can drastically change the quality of recovered parameters for a given microstructure descriptor. When assessing the invertibility of different manifold construction techniques, performance is not just a function of descriptor choice in the forward process $f(\Theta)$, but also regression model choice in the inverse process ($g(f(\Theta)$). Some descriptors may excel for a linear regression model, while others may perform better using non-linear approaches.

% \textbf{Full-Cycle Invertibility Example}

% Cyclic invertibility is needed due to the highly iterative nature of materials development. If the inverse model is not able to accurately map between domain and co-domain, sequential iterations will begin to fall off of their target microstructures, leading to poor optimization in the exploration of the material design space. Figure \ref{fig:cycled_inverse} shows an example of what one full iterative loop looks like, beginning with input parameters: $\theta=$($\eta=0.04$, $\kappa=0.58$), and the micrograph generated from those parameters in \ref{fig:cycled_inverse}(a). From there, the $D_{PH}$ descriptor vector was calculated on that image and passed through the LinearSVR regression model to recover its predicted parameters: $\theta'=$($\eta=0.02$, $\kappa=0.57$). Those parameters were passed back into the phase field generator (with the same random seed, $\xi$) to generate the micrograph shown in Figure \ref{fig:cycled_inverse}(b). While there are some small changes in connectivity in the image, on the whole, the regenerated microstructure is extremely close to the input microstructure, beyond a practitioner's ability to differentiate between $M$-states.

% \begin{figure}[h!] \begin{centering} \includegraphics[width=0.75\linewidth]{cycled_inverse.png} \caption{Example microstructure and its full-cycled comparison.} \label{fig:cycled_inverse} \end{centering} \end{figure}

\subsection{Manifold Visualization} \label{sec:mani_vis}

Lower-dimensional projections ($\mathbb{R}^2/\mathbb{R}^3$) of the higher-dimensional microstructure descriptor spaces were used to qualitatively explore the approximation of the material manifold. A general assessment of the coherency and navigability of the manifolds can be done by looking for continuity between the parameter-based projection of the manifold and those constructed based on the microstructure descriptors.

For each of the manifold projection techniques, the family of $\hat{M}$-approximations for all $\theta$-parameter sets in the material system were used to find low-dimensional projections of the manifold.

Principal component analysis is a linear dimensionality reduction technique that transforms higher-dimensional data into a lower dimension, calculating the linear combination of variables in the original representation that explain the greatest variance in the system \cite{wold1987principal}. It has previously been used for exploring material systems \cite{niezgodaUnderstandingVisualizingMicrostructure2011} and is a quick method for exploring linear relationships in the embedded microstructure domain.

t-distributed stochastic neighbor embedding (tSNE) was chosen as a manifold projection technique as it constructs a non-linear manifold based on the local similarities between data points \cite{van2008visualizing}. This was appealing for our method, as the pre-calculated MDDs between $\hat{M}^\theta$-approximations can be used to construct the nearest-neighbor relationships, and the `perplexity' parameter can be tuned to adjust the degree of locality versus globality of the projection.

These manifold projection techniques were calculated using the Scikit-Learn's `PCA' and `TSNE' functions \cite{scikit-learn}. The data was taken from the embedded microstructure domain ($\hat{M} \in \mathbb{R}^{100}$ or $\hat{M} \in \mathbb{R}^{200}$) down to a lower projected domain in $\mathbb{R}^2$ or $\mathbb{R}^3$. The dimensionality of the projected domain was chosen in conjunction with the results from the dimensionality estimation in Section \ref{sec:dimEst}, as the minimum dimensionality of the projected domain for an interpretable visualization should match the estimated dimensionality, $\mathbb{R}^{estimated}$, or $\mathbb{R}^{estimated+1}$ in the case of a surface embedded in a higher dimensional domain.

\begin{figure}[h!] \begin{centering} \includegraphics[width=1.0\linewidth]{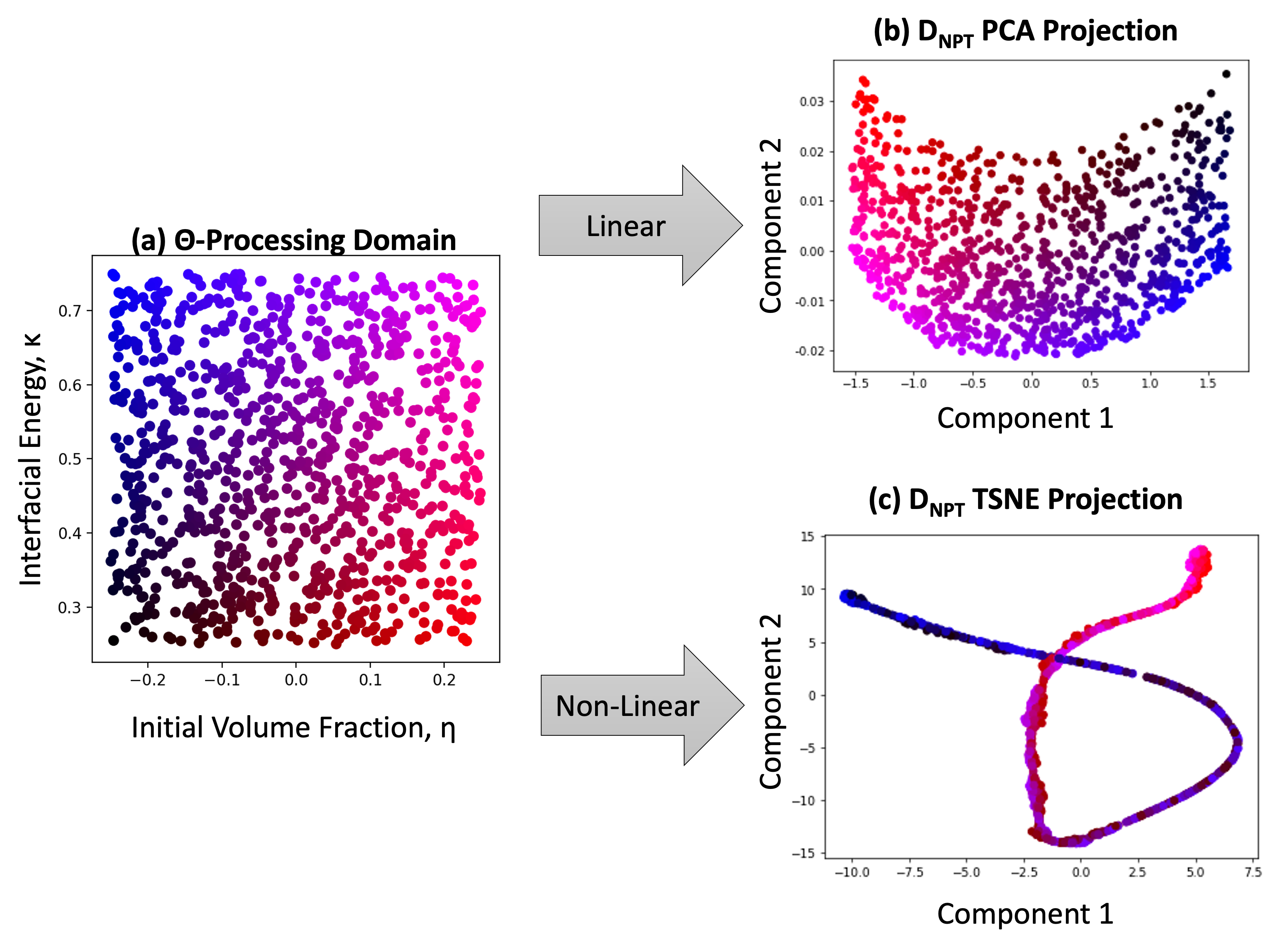} \caption{(b) PCA and (c) tSNE projections of $\hat{M}$-approximations of the $\mathcal{M}$-manifold using $D_{NPT}$} \label{fig:npt_proj} \end{centering} \end{figure}

Figure \ref{fig:npt_proj} shows manifold projections of the $\hat{M}$-approximations using $D_{NPT}$ for the (b) PCA and (c) tSNE techniques respectively. These plots use the same red/blue color scheme established in Section \ref{sec:datasetSampling} to qualitatively interpret how cohesive these manifold approximations are by looking for smooth color gradients across the surfaces in comparison to Figure \ref{fig:npt_proj}(a). The PCA projection shown in Figure \ref{fig:npt_proj}(b) is visually similar to that of the $\Theta$ domain in (a); a 2D plane with a slight bend to it. It is important to note the change in scale from Component 1 to Component 2 in Figure \ref{fig:npt_proj}(b), where Component 1 is two orders of magnitude greater than that of Component 2. In contrast, the tSNE projection shown in \ref{fig:npt_proj}(c) appears as a thin ribbon or 1D line embedded in 2D space, unable to separate out the $\kappa$ parameter. These results, paired with the analysis from Section \ref{sec:invert} give rise to the hypothesis in Section \ref{sec:invert} that the MDD values calculated using $D_{NPT}$ are skewed against the $\kappa$ parameter.

\begin{figure}[h!] \begin{centering} \includegraphics[width=1.0\linewidth]{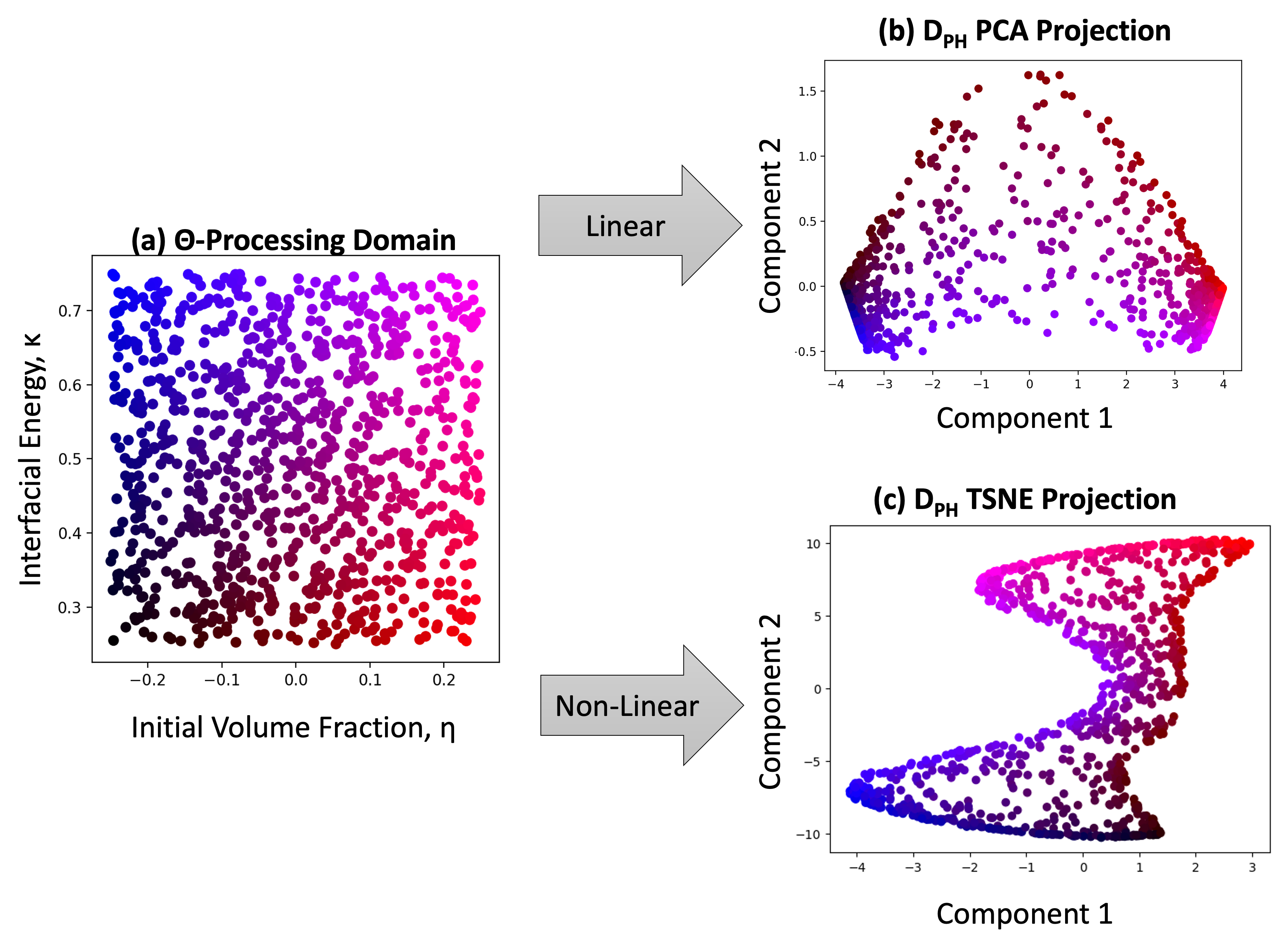} \caption{(b) PCA and (c) tSNE projections of $\hat{M}$-approximations of the $\mathcal{M}$-manifold using $D_{PH}$} \label{fig:ph_proj} \end{centering} \end{figure}

Figure \ref{fig:ph_proj} shows the PCA and tSNE manifold projections of the $\hat{M}$-approximations using $D_{PH}$. In contrast to $D_{NPT}$, the PCA projection shows more artifacts of the descriptor choice and the tSNE projection gives a more cohesive manifold projection overall. Again looking like a `bent' version of the processing domain in Figure \ref{fig:ph_proj}(a), the 2D PCA projection in Figure \ref{fig:ph_proj}(b) shows an increase in sparsity in the intermediate regime of the manifold, having densely-packed points of $M$-states along the two extreme edges (in terms of $\eta$). This artifact comes about due to the increased rate of change of $D_{PH}$ regarding $\Delta\theta$ in the intermediate, highly-connected regime of the parameter space ($\eta$ between approx. -0.05 and 0.05). This region on the $\mathcal{M}$-manifold is seen as ``unstable'' regarding the MDD calculations on the $D_{PH}$ as the persistent homology descriptor is sensitive to changes in system topology and connectivity as the $M$-states undergo a phase inversion. The tSNE projection has more evenly distributed $M$-states in the structure domain, while having a few interesting characteristics described ahead in Section \ref{sec:mapping}. As tSNE has the perplexity parameter to tune the local/global nature of the projection, it is able to compress and decompress the various non-linear regions of the domain.

\begin{figure}[h!] \begin{centering} \includegraphics[width=1.0\linewidth]{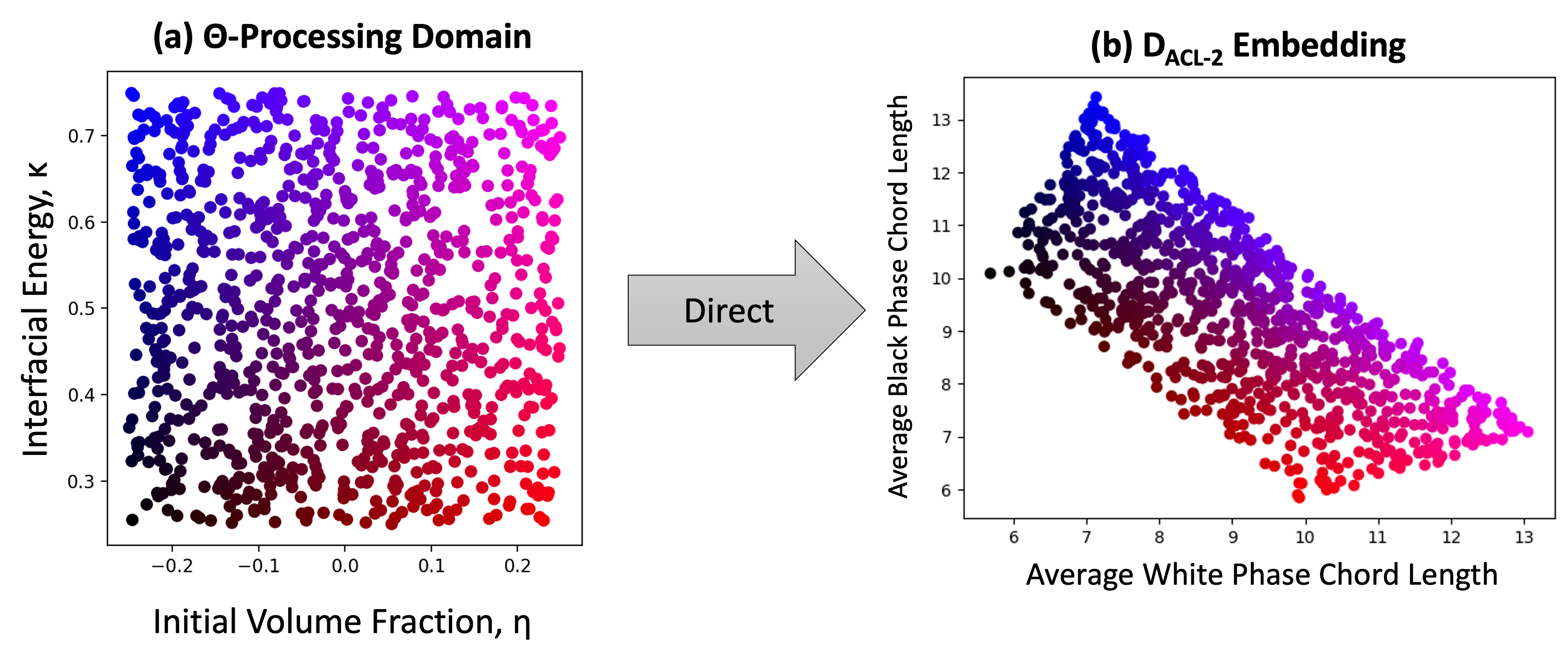} \caption{(b) 2D embedding of $\hat{M}$-approximation of the $\mathcal{M}$-manifold using $D_{ACL-2}$} \label{fig:acl_proj} \end{centering} \end{figure}

Figure \ref{fig:acl_proj} plots the $M$-states in the embedded $D_{ACL-2}$ domain, without the need for dimensionality reduction via PCA or tSNE, as the descriptor natively calculates a latent domain of 2 dimensions. Figure \ref{fig:acl_proj}(b) shows this embedding with the x-axis corresponding to the average chord length of the white phase, and the y-axis corresponding to the average chord length of the black phase. Following the prior dimensionality and invertibility assessments, this projection of the data validates those results, establishing a cohesive latent representation of the $M$-states and allowing for navigation. While it remains to be seen if this simple descriptor in $\mathbb{R}^2$ can be used for forward analysis, for example in material property prediction, it provides one great benefit in terms of navigability and interpretability in that each of the latent domains relates directly to a physical microstructural feature.

\section{Discussion}
\subsection{Mapping Microstructure: From Processing to Structure Domains} \label{sec:mapping}

\begin{figure}[!ht] \begin{centering} \includegraphics[width=1.0\linewidth]{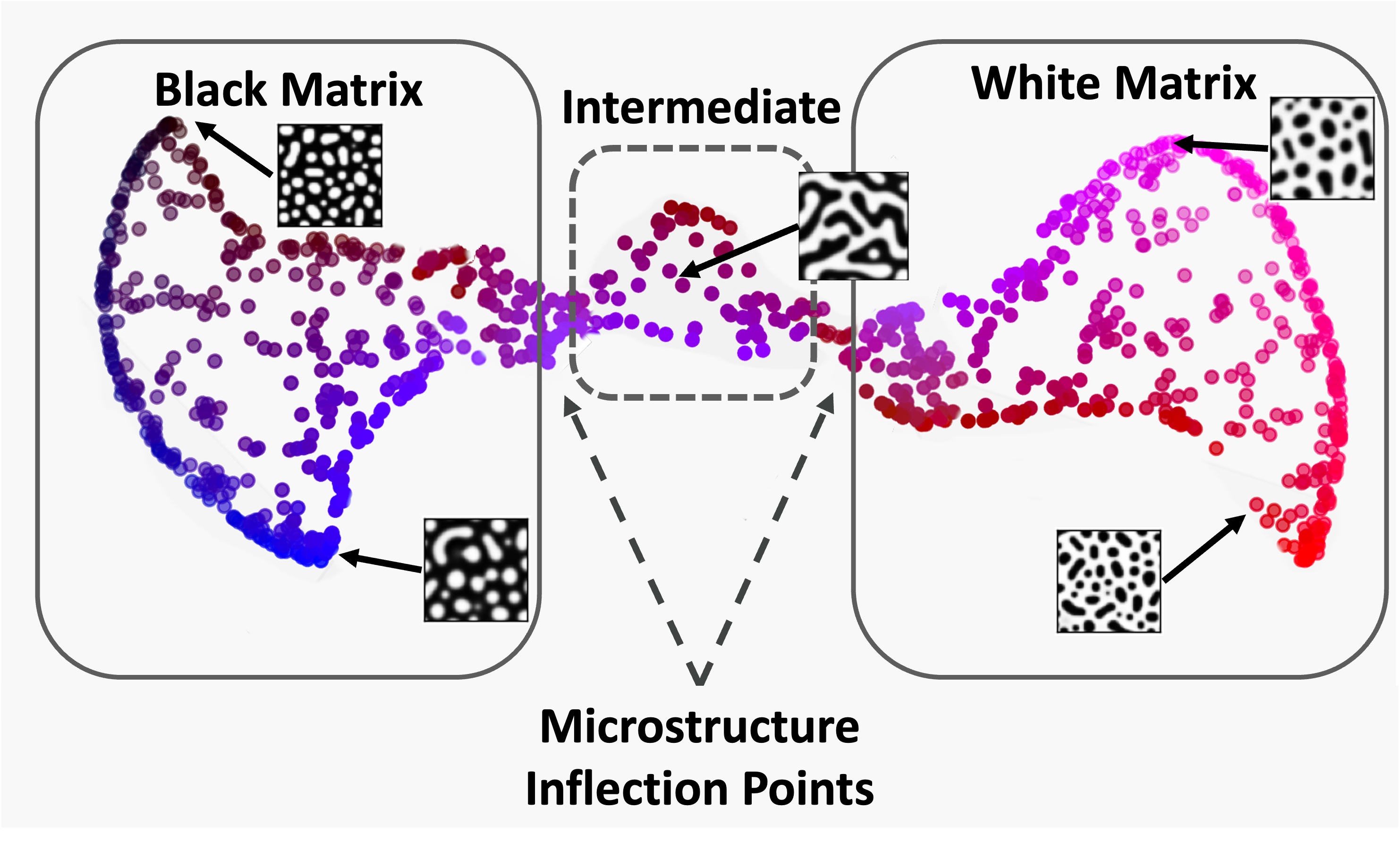} \caption{2D unraveling of tNSE projection of $D_{PH}$.} \label{fig:ph_tsne_annotated_proj} \end{centering} \end{figure}

Building from the results in the Sections \ref{sec:dimEst} and \ref{sec:invert} and visualizations in Section \ref{sec:mani_vis}, different approaches to summarizing salient microstructural information provide alternative interpretations of the underlying domain, although no single approach is inherently incorrect or better than the others. Revisiting Figure \ref{fig:ph_proj}(c) with a slightly adjusted perplexity parameter for easier visualization, Figure \ref{fig:ph_tsne_annotated_proj} shows a tSNE projection of the $D_{PH}$ representation of the latent domain. This visualization technique highlights the natural separation of the $M$-states into three different regions along the structure manifold: (1) the primary black region on the left, (2) the intermediate region in the middle, and (3) the primary white region on the right. With this non-linear representation of the $\mathcal{M}$-manifold, it becomes immediately apparent how to navigate the manifold in such a way to either retain a microstructure having a certain topology, or to traverse the manifold and enter a new regime of connectivity. On the other hand, $D_{NPT}$ and $D_{ACL-2}$ provide much more uniform representations of the data, with consistent steps from one $M$-state to the other, excelling at extracting a linear relationship of the underlying $\mathcal{M}$-manifold.

\subsection{Attributes of an ``Informative'' Microstructure Representation} \label{sec:attributes}

\begin{figure}[!h] \begin{centering} \includegraphics[width=0.8\linewidth]{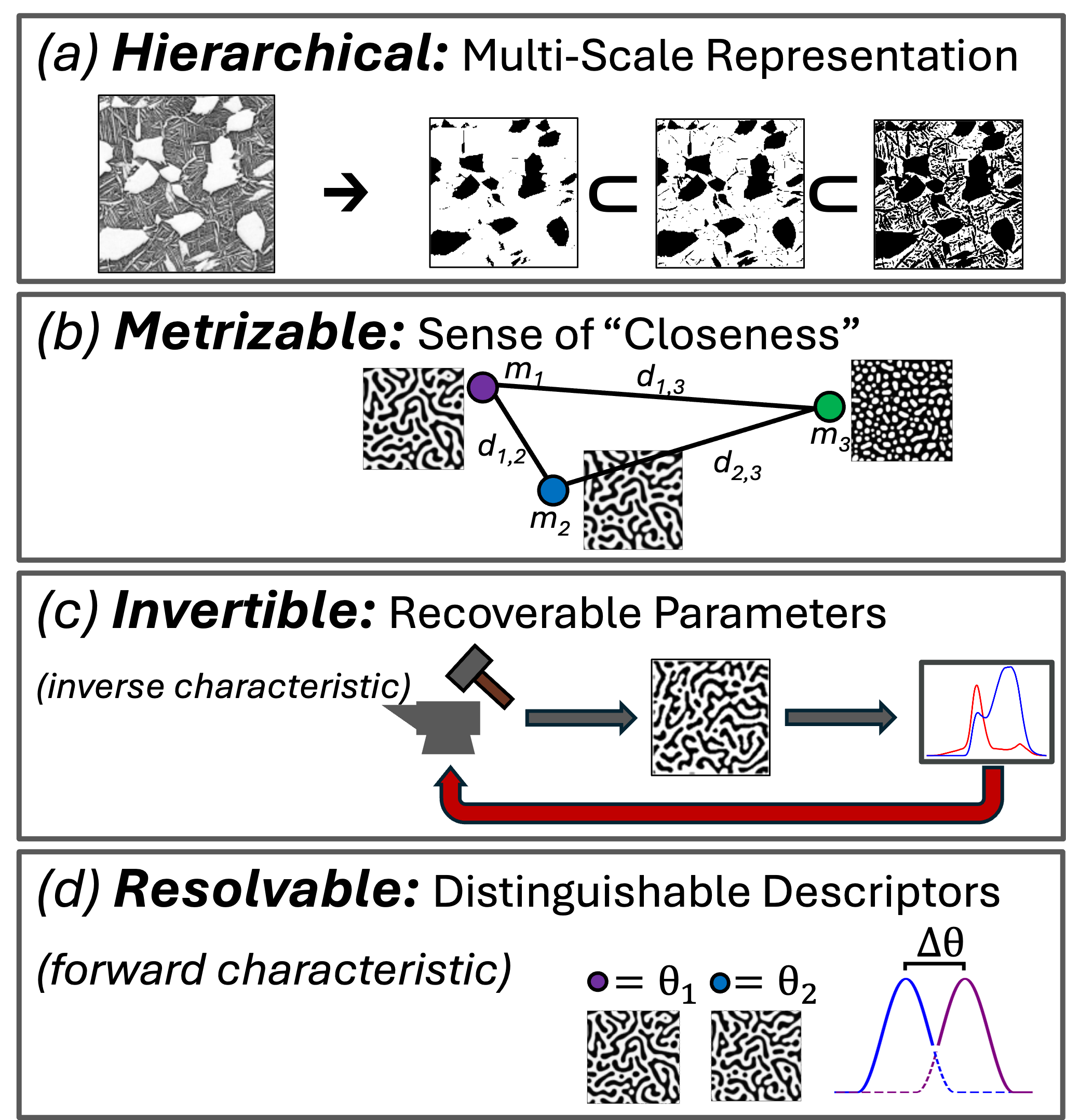} \caption{Attributes of an ``informative'' microstructure representation.} \label{fig:attributes} \end{centering} \end{figure}

In search of the $\mathcal{M}$-manifold, we have determined necessary criteria for microstructure descriptors to be conducive to the construction of a continuous space. The attributes required are shown in Figure \ref{fig:attributes} and outlined below:

\begin{enumerate}

\item \textbf{Hierarchical:} Microstructure is inherently a multi-scale problem. Being able to quantify features spanning across multiple length scales and understand the effects of both short- and long-range order are key to having a robust understanding of microstructure.

\item \textbf{Metrizable:} Having a sense of closeness is crucial for defining the neighborhood relationships between $M$-states. Strictly, metrizability requires the Big $M$ domain ($\mathcal{M}$-Manifold) to exist on a metric space with the ability to calculate true distance measures between $M$-states.

\item \textbf{Invertible:} Invertibility is the ability of a model to recover the original input parameters from the descriptor space.

\item \textbf{Resolvable:} Resolvability is the notion of having distinguishable descriptors, or being able to recognize different $M$-states within some given parameter-threshold. 

\end{enumerate}

The majority of this work focused on the \textbf{metrizability} characteristic, illustrating its importance in establishing local, regional, and global representations of the material manifold. Subsequent analysis methods were built on top of the quantitative descriptor metrics. 

Initial assessment has been performed on the \textbf{invertibility} requirement of an informative microstructure representation, but only in a manifold-averaged sense. Further development on the invertibility of different descriptor choices is needed to understand how invertibility varies across the span of the manifold.

Going back to Section \ref{sec:funcProc} on treating microstructure as a function of processing, invertibility is a characteristic of the inverse process: i.e. the mapping from microstructure space back to processing space $g: M \rightarrow \Theta$. While this is an important notion in establishing an informative feedback loop, it represents half of the problem. \textbf{Resolvability} acts as its foil in understanding the model variability as a function of the forward process, meaning that it is defined by the mapping from processing space to microstructure space $f:\Theta \rightarrow M$ (or more correctly, to microstructure descriptor space). For a successful feedback loop to occur, both the forward and inverse processes must be well-established. If the forward process is not well-defined (e.g. a poor choice of microstructure descriptor), the inverse process will not be able to recover meaningful information. Work on the establishment of resolvability limits is ongoing.

\subsection{Potential Application to Experimental Data}

The benefits achieved by manifold-based materials exploration are complimentary to both traditional and accelerated materials development techniques. While experimental data collection is limited in the number of data points that can be collected, either in terms of images ($m$-instances) or samples ($M$-states), the manifold approach can alleviate some of these issues. The stochastic approximation can increase the robustness of descriptors measured on noisy datasets. Resource limitations in terms of data collection can be addressed when integrating the manifold approach with accelerated materials development techniques, like active learning \cite{moritaApplyingBayesianOptimization2022}, \cite{liuActiveLearningRegression2024}. These methods allow for on-the-fly decision-making with respect to determining when enough individual realizations have been captured for a given material state, as well as on determining which material state to explore next \cite{kusneOntheflyClosedloopMaterials2020}. Harnessing the combined utility of these novel materials development strategies has the potential to shorten the number of development iterations needed to take new materials from discovery to deployment.

%%%%%%%%%%%%%%%%%%%%%%%%%%%%%%%%%%%%%%%%%%%%%%%%%%%%%%%%%%%% %%%%%%%%%%%%%%%%%%%%%%%%%%%%%%%%%%%%%%%%%%%%%%%%%%%%%%%%%%%%

\section{Conclusions}

In this study we introduced and evaluated a data‑driven framework for constructing a material manifold --- a low--dimensional latent space in which each point uniquely represents a material state defined by its processing conditions. The central tenet of the approach is to treat microstructure not as a single, deterministic image but as a stochastic process from which many independent $m$‑instances can be sampled. By extracting distribution--based descriptors from ensembles of $m$‑instances, we obtain quantitative material--state vectors that are both metrizable and amenable to manifold learning. Distances between material states reflect true differences in the \textit{effects} of processing parameters, enabling reliable dimensionality estimation.

Applying this approach to phase--field simulations of spinodal decomposition demonstrates that descriptors such as persistent homology, two--point statistics, and average chord length successfully recover the intrinsic two--dimensional nature of the system, whereas direct image distances fail when stochastic variability is introduced. Invertibility is confirmed through regression analyses that retrieve the original processing variables, showing that the forward mapping from process to descriptor space and the inverse mapping back are both well conditioned. Visualizations using PCA and tSNE illustrate a smooth surface that mirrors the underlying parameter domain, reinforcing the notion of continuity, where small changes in processing lead to small, predictable changes in microstructure.

By re--conceptualizing microstructure as a stochastic ensemble and leveraging distribution--based descriptors, we have demonstrated that complex processing--microstructure relationships can be collapsed onto an interpretable, low--dimensional material manifold. The manifold is both locally continuous, ensuring smooth navigation across the design space, and invertible, enabling reliable recovery of processing parameters from microstructural measurements. This quantitative foundation opens a pathway towards accelerated materials discovery: rapid generation of candidate process recipes, systematic exploration of the state space, and closed‑loop optimization of target properties --- all within a unified, data--centric framework.

%%%%%%%%%%%%%%%%%%%%%%%%%%%%%%%%%%%%%%%%%%%%%%%%%%%%%%%%%%%%
%%%%%%%%%%%%%%%%%%%%%%%%%%%%%%%%%%%%%%%%%%%%%%%%%%%%%%%%%%%%

\section{Acknowledgments}

The authors would like to thank the range of funding opportunities that enabled this collaborative effort: Much of this work was completed as part of S.A. Mason's thesis work at The Ohio State University. S.A. Mason and S.R. Niezgoda received funding in part from the National Science Foundation's ERC-HAMMER program under NSF CMMI 2133630. This research was supported in part by the Air Force Research Laboratory Materials and Manufacturing Directorate, through the Air Force Office of Scientific Research Summer Faculty Fellowship Program, Contract Numbers FA8750-15-3-6003, FA9550-15-0001 and FA9550-20-F-0005, working in collaboration with M.N. Shah and J.P. Simmons under AFRL-2025-3988. D.M. Dimiduk acknowledges partial support from BlueQuartz Software, LLC. Additionally, the authors would like to thank Alan Gan and Prof. Oksana Chkrebtii from The Ohio State University's Department of Statistics for their guidance and and feedback throughout this work.

\section{Declarations}

\textbf{Conflict of Interest:} On behalf of all authors, the corresponding author states that there is no conflict of interest.

\textbf{Data Availability:} Data and code are openly available on Zenodo at \url{https://doi.org/10.5281/zenodo.17456299} \cite{mason2025-dataset}.

%%%%%%%%%%%%%%%%%%%%%%%%%%%%%%%%%%%%%%%%%%%%%%%%%%%%%%%%%%%%
%%%%%%%%%%%%%%%%%%%%%%%%%%%%%%%%%%%%%%%%%%%%%%%%%%%%%%%%%%%%

\newpage

\singlespacing

\section{References}
\renewcommand\refname{}
\bibliography{main}
% \printbibliography[heading=none]

\end{document}